\documentclass[10pt,journal,compsoc]{IEEEtran}
\usepackage{graphicx}
\usepackage{listings}
\usepackage{xcolor}
\usepackage{amsmath}
\usepackage{booktabs}
\usepackage{tabularx}
\usepackage{makecell}
\usepackage{cite}
\usepackage{multirow}
\usepackage{url}
\begin{document}

%
\title{SolidityCheck : Quickly Detecting Smart Contract Problems Through Regular Expressions}
\author{Pengcheng~Zhang, \emph{Member, IEEE},
        Feng Xiao,
        Xiapu Luo
\IEEEcompsocitemizethanks{\IEEEcompsocthanksitem P. Zhang and F.Xiao are with the College of Computer and Information,
Hohai University, Nanjing, P.R.China\protect\\
E-mail: pchzhang@hhu.edu.cn; harleyxiao@foxmail.com
\IEEEcompsocthanksitem X.Luo is with the Department of Computing, the Hong Kong Polytechnic University, HongKong, China\protect\\
E-mail:csxluo@comp.polyu.edu.hk

}
\thanks{Manuscript received XXXX XXXX; revised XXXX, XXXX.}}

\IEEEtitleabstractindextext{
\begin{abstract}
  As a blockchain platform that has developed vigorously in recent years, Ethereum is different from Bitcoin in that it introduces smart contracts into blockchain.
  \emph{Solidity} is one of the most mature and widely used smart contract programming language, which is used to write smart contracts and deploy them on blockchain. However, once the data in the blockchain is written, it cannot be modified. Ethereum smart contract is stored in the block chain, which makes the smart contract can no longer repair the code problems such as re-entrancy vulnerabilities or integer overflow problems.
  Currently, there still lacks of an efficient and effective approach for detecting these problems in \emph{Solidity}. In this paper, we first classify all the possible problems in \emph{Solidity}, then propose a smart contract problem detection approach for \emph{Solidity}, namely \emph{SolidityCheck}.
 The approach uses regular expressions to define the characteristics of problematic statements and uses regular matching and program instrumentation to prevent or detect problems. Finally, a large number of experiments is performed to show that \emph{SolidityCheck} is superior to existing approaches. 
\end{abstract}
}

\maketitle

\IEEEdisplaynontitleabstractindextext

%
\IEEEpeerreviewmaketitle

\section{Introduction}\label{sec:introduction}

Ethereum is the largest blockchain that supports smart contracts with a market capital of 18 billion~\cite{wood2014ethereum}. Smart contracts~\cite{luu2016making} are autonomous programs running on the blockchain platform. They are usually developed in several high-level languages and then compiled into bytecode. Once the bytecode of smart contract is deployed to blockchain, its functions can be invoked by others but the bytecode cannot be changed. Unfortunately, it is inevitable that many smart contracts contain bugs but they cannot be patched because of the data immutability of blockchain~\cite{tikhomirov2018smartcheck,nikolic2018finding}. 
Consequently, it is particularly important to have automated tools that can help developers thoroughly check their smart contracts before deploying their bytecode to the blockchain. 



\begin{table}[h]
\centering
\caption{Classification definition of smart contract problems}\label{Definition}
\begin{tabular*}{9cm}{p{2.5cm}<{\centering}|p{4.3cm}<{\centering}|c}
\toprule
Types of problems & Problem Description & Severity\\
\midrule
Security problem & Vulnerabilities in contract code cause developers to suffer losses & High\\
\midrule
Performance problem & Smart contract execution costs too much or performs poorly due to the use of certain statements & Medium\\
\midrule
Hidden threats of coding problems & Statements that may cause security problems in specific situations or reduce code readability &  Low\\
\bottomrule
\end{tabular*}
\end{table}

A number of recent studies report the possible issues in smart contracts~\cite{kalra2018zeus,krupp2018teether,torres2018osiris,fontein2018comparison,grishchenko2018foundations,parizi2018empirical}. Base on existing work, we classify them into three categories as listed in Table~\ref{Definition}.

\emph{Security problems}. Vulnerabilities in smart contract codes cause developers to suffer losses. For example, DAO, the largest crowdsourcing project in Ethereum, was found to have a re-entrancy vulnerability in its code, resulting in the loss of \$12 million worth of ethers in 2016~\cite{mehar2019understanding}.




\emph{Performance problems.} This kind of problems increases gas consumption for running contracts. Chen et al.~\cite{chen2018understanding} analyzed the deployed smart contracts and found that more than 80\% of the smart contracts have performance related problems, even though the codes have been optimized by the recommended compiler.

\emph{Hidden threats of coding problems}. Statements that may cause penitential security problems in specific situations or reduce code readability are defined hidden threats of coding problems. In general, these problems do not necessarily cause serious problems. However, paying attention to these problems and solving them can make smart contracts safer and easier to maintain.


Some tools have been proposed to check the problems aforementioned in smart contracts~\cite{grishchenko2018semantic,luu2016making,albert2019safevm,tann2018towards,nikolic2018finding,torres2018osiris,tsankov2018securify,chen20171,bragagnolo2018smartinspect}. However, most of them can only handle the bytecode of smart contracts. Although processing bytecode directly empowers the tools to analyze all deployed smart contracts, they cannot leverage the useful information in source codes (e.g., naming functions and events) and consequently a tool that can quickly and accurately locate the issues in the source codes of smart contracts would be more useful for the smart contract developers who have the source codes at hand. Moreover, our experiments in Section~\ref{exp_efficiency} and other studies~\cite{luu2016making,tsankov2018securify,nikolic2018finding,torres2018osiris} show that the bytecode based tools are less efficient than tools for handling source codes. Although a recent work~\cite{tikhomirov2018smartcheck} designed a tool (named \emph{SmartCheck}) for finding problems from the source codes of smart contracts, existing work has the following limitations: 
\begin{enumerate}
\item Existing vulnerability detection criteria is confusing. They do not accurately characterize some problems. For example, an external function call followed by an internal function call is identified as having a re-entrancy vulnerability, which could not accurately capture the characteristics of re-entrancy vulnerability and would cause a large number of misjudgments and omissions.
\item \emph{SmartCheck} cannot detect some important security problems, such as integer overflow. 
Missing this problem may lead to serious consequences. For example, the integer overflow problem led to the big loss of the \emph{BEC project}.

\item The detection efficiency of the existing work is very low. \emph{SmartCheck} runs lexical and grammatical analysis on the \emph{Solidity} source codes and then generates the corresponding XML parse tree for the source codes. Based on the parse tree, it uses XPath to retrieve the problematic statement~\cite{tikhomirov2018smartcheck}. Lexical analysis and grammatical analysis reduce the efficiency of \emph{SmartCheck} analysis.
\end{enumerate}


To address these limitations, in this paper, we propose, \emph{SolidityCheck}, a novel approach using regular expressions to quickly and accurately locate 20 kinds of problems in the source codes of smart contracts. In particular, we can prevent two particularly dangerous security problems: (re-entrancy and integer overflow). We also conduct extensive experiments to evaluate the usability, efficiency and effectiveness. In summary, we make the following novel contributions: 


\begin{itemize}
\item We propose a new classification criterion, which identifies 20 kinds of code problems that have adverse effects on smart contracts, including several previously undetected ones, covering the vast majority of smart contract problems currently.
\item We detect problematic statements through regular expressions. Most kinds of problematic statements can be accurately detected by regular expressions, and then the test results can be reported to the user for easy modification. Furthermore, two important problems which are difficult to detect and have a significant impact on the security of smart contracts are retrieved by regular expressions, and then these problems are prevented by program instrumentation.
\item  We particularly designed a set of experiments to validate \emph{SolidityCheck}. The experimental results show that our tool is superior to the existing static code analysis tools in recall, precision, and other indexes.
\end{itemize}

The rest of this paper is organized as follows. Section~\ref{sec_Preliminaries} provides the basic concepts used in this paper. Section~\ref{sec_class} proposes a novel classification criterion for smart contract problems and discusses the characteristics of each problem. We detail the design and the implementation of \emph{SolidityCheck} in Section~\ref{Sec_Soliditycheck}. 
Section~\ref{Sec_Experimental} reports our extensive evaluation results of \emph{SolidityCheck}. We discuss the limitations of \emph{SolidityCheck} in Section~\ref{sec_disucssion}. After discussing the related work in Section~\ref{sec_Related work}, we conclude the paper and point out the future work in Section~\ref{sec_conclusion}.

\section{Preliminaries}\label{sec_Preliminaries}

\subsection{Smart contract}
Smart contracts are computer programs that can automatically execute contract terms~\cite{he2017survey}. Smart contracts are automatically executed when their execution conditions are satisfied, and the execution results are generated according to the behaviors in the contracts. Using a smart contract to sign a contract can effectively avoid disputes. Blockchain is well suitable for the operating environment of smart contracts because of its decentralization and network-wide consensus.
Ethereum smart contract accounts share the same address space with the external accounts, and the smart contract can be invoked by sending transactions to the contract address. To prevent the unwarranted waste of Ethereum's calculating power, Ethereum collects gas from each executed smart contract statement, which is converted from ethers.
\subsection{Solidity}
\emph{Solidity} is the most mainstream, mature and widely used Ethereum smart contract programming language~\cite{atzei2017survey}. Unlike the lower-level language, \emph{Solidity} is a Turing-complete high-level programming language, which is capable of expressing arbitrary complex logic. 
Smart contracts programmed in \emph{Solidity} language are compiled into the Ethereum virtual machine bytecodes and running in each Ethereum node. 
\emph{Solidity}language is specially developed for the compilation of Ethereum smart contracts. It contains built-in functions to complete various functions of Ethereum. For example, \emph{transfer} and \emph{send} functions are used to execute transfer ethers, and keywords such as \emph{require} and \emph{assert} are designed for checking status. \emph{Solidity} is a fast iterative language. The same keyword may have different semantics in different language versions. To improve this situation, when smart contracts in \emph{Solidity} are written, it is necessary to specify the versions of the compiler that the contract can accept.
\section{Classification of Existing Problems in Smart Contracts}   \label{sec_class}
\begin{figure*}[h]
    \centering
    \includegraphics[scale=0.3]{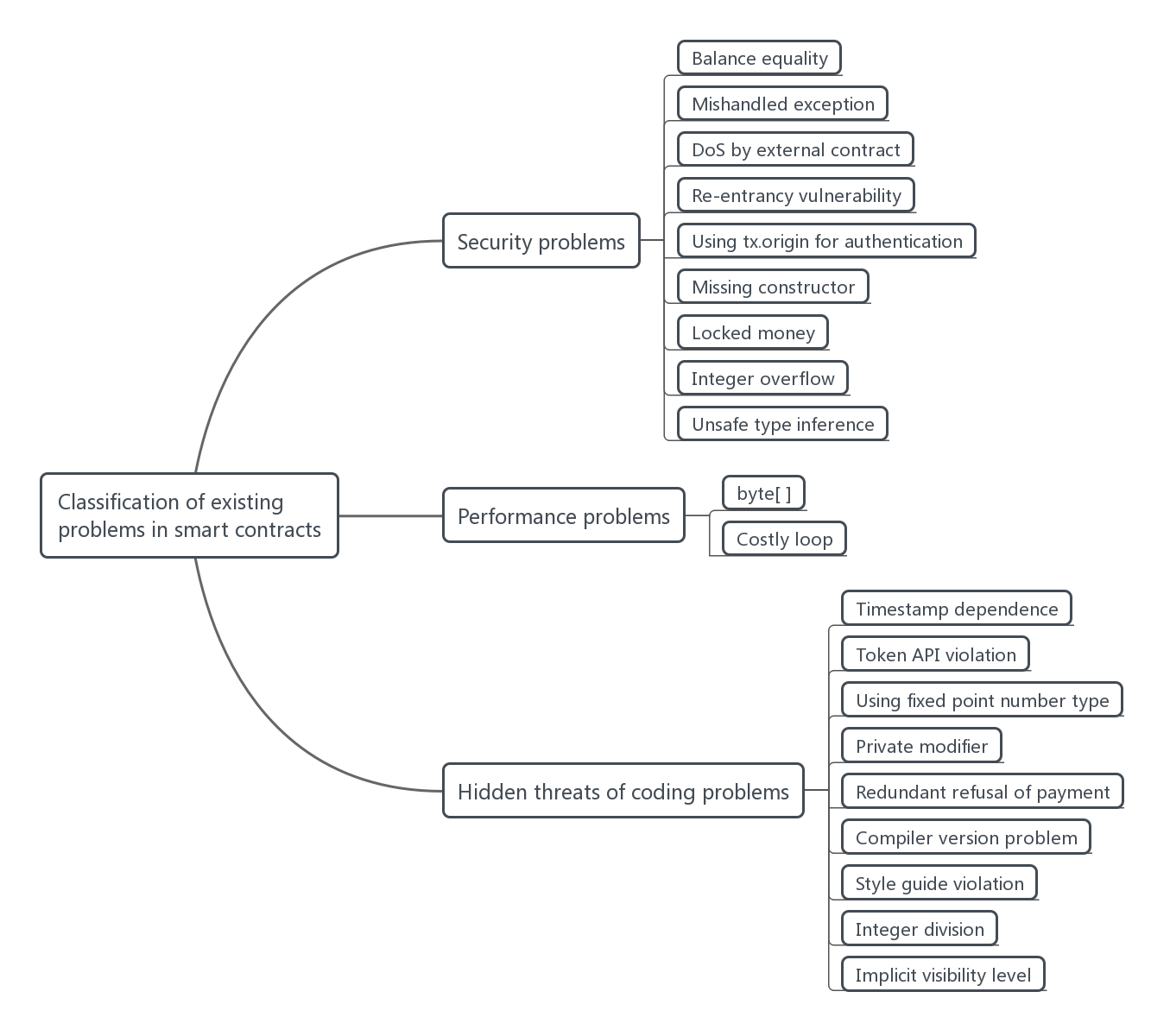}
    \caption{A classification criterion for smart contract problems}
    \label{fig_claffication}
    \end{figure*}

Based on existing studies~\cite{tikhomirov2018smartcheck,nikolic2018finding,kalra2018zeus,krupp2018teether,torres2018osiris,fontein2018comparison,grishchenko2018foundations,parizi2018empirical}, we give a classification criterion for smart contract problematic statements (see Fig.~\ref{fig_claffication} for detail), which summarizes 20 kinds of common problems that need to be detected.
Furthermore, we also describe the consequences of each problem and the corresponding detection approach we propose. The corresponding regular expressions we designed for these detection approaches are shown in appendix A. 



\subsection{Security Problems}

\noindent\emph{Balance equality}~\cite{tikhomirov2018smartcheck}. 
An adversary can forcibly send ethers to the attacked contract by mining or via \emph{selfdestruct}, so that the conditional judgment part in Listing 1 is always false.
           \begin{lstlisting}[     language=C++,
                            breaklines=true,
                            captionpos=bc,
                            basicstyle = \footnotesize,
                            title={Listing 1: balance equality},
                            keywordstyle=\color{red},
                            commentstyle=\color{blue},
                            stepnumber=1
                            ]
    if (this.balance == 1997 ether){
        //do something
    }
    \end{lstlisting}

\noindent\emph{Mishandled exception}~\cite{luu2016making}. In Ethereum, contracts can call other contracts in several ways (eg., via \emph{send}, \emph{delegatecall} or \emph{call}~\cite{krupp2018teether}). If an exception occurs in the callee contract, the call terminates, rolls back the status of the callee contract and returns false. Therefore, the return value of an external call should be checked to properly handle the exception~\cite{kalra2018zeus,luu2016making}. Listing 2 shows a possible loss scenario in which the contract reduces \textbf{addr}'s holding of tokens when \textbf{addr} fails to receive the transfer for some reasons.
          \begin{lstlisting}[     language=C++,
                            breaklines=true,
                            captionpos=bc,
                            basicstyle = \footnotesize,
                            title={Listing 2: mishandled exceptions},
                            keywordstyle=\color{red},
                            commentstyle=\color{blue},
                            stepnumber=1
                            ]
    addr.call.value(1 wei); //transfer 1 wei to addr
    balance[addr] -= 1;     //reduce addr's tokens
    \end{lstlisting}

\noindent\emph{DoS by external contract}~\cite{tikhomirov2018smartcheck}.
External contracts may be maliciously controlled or killed, which may result in the invalidation of some or all functions of this contract. As shown in Listing 3, when the dependent external contract self-destructs, the function \emph{getService} fails. It is particularly noteworthy that when contracts depend on external libraries, the security of libraries should be carefully reviewed. 
          \begin{lstlisting}[     language=C++,
                            breaklines=true,
                            captionpos=bc,
                            basicstyle = \footnotesize,
                            title={Listing 3: dos by external contract},
                            keywordstyle=\color{red},
                            commentstyle=\color{blue},
                            stepnumber=1
                            ]
  function getService(address _provider, address _customer) public{
    Provider provider = Provider(_provider);
    //if _customer is a user of the service,
    //the _customer can get service of the _provider
    if(provider.isCustomer(_customer)){
        //providing service
    }
  }
    \end{lstlisting}


\noindent\emph{Re-entrancy vulnerability}~\cite{tsankov2018securify}. The re-entrancy vulnerability leads to the dissolution of The DAO and the division of the Ethereum community. 
Source based code analysis cannot accurately determine whether a statement or a piece of codes introduces re-entrancy vulnerabilities. 
 Listing 4 shows a contract with a re-entrancy vulnerability (hereinafter referred to as an attacked contract). An attacker can write a specific attack contract. He can first deposit ethers into the attacked contract through the attack contract, and then retrieve his deposit by calling \emph{withdrawBalance} function in the attacked contract. The \emph{fallback} function of the attacking contract calls again the \emph{withdrawBalance} function of the attacked contract, but now the balance of the attacking contract has not deductions, consequently it can withdraw many times.
          \begin{lstlisting}[     language=C++,
                            breaklines=true,
                            captionpos=bc,
                            basicstyle = \footnotesize,
                            title={Listing 4: an smart contract with re-entrancy vulnerability},
                            keywordstyle=\color{red},
                            commentstyle=\color{blue},
                            stepnumber=1
                            ]
pragma solidity ^0.4.15;
    
contract Reentrance{
    mapping (address => uint) userBalance;
        
    function withdrawBalance(){
        //send userBalance[msg.sender] ethers to msg.sender
        //if msg.sender is a contract, it responds with the fallback function.
        if (msg.sender.call.value(userBalance[msg.sender])()){
              throw;
        }
        userBalance[msg.sender] = 0;
    }
}
\end{lstlisting}

\begin{figure}[h]
    \centering
    \includegraphics[scale=0.34]{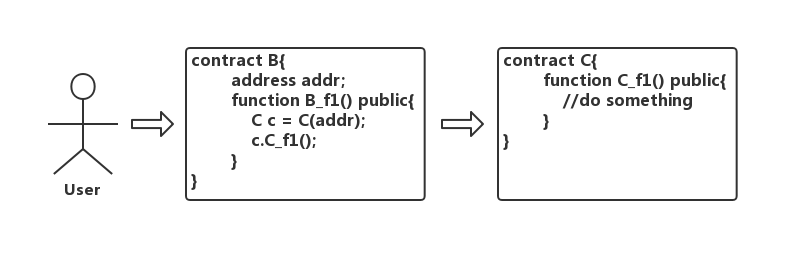}
    \caption{Differences between tx.origin and msg.sender}
    \label{fig7}
    \end{figure}

\noindent \emph{Using \emph{tx.origin} for authentication}~\cite{tikhomirov2018smartcheck}. \emph{tx.origin} is different from \emph{msg.sender} (both keywords are provided in Solidity). \emph{tx.origin} points to the initiator of the transaction, while \emph{msg.sender} is the sender of the message. \emph{tx.origin} always points to the external account controlled by the user. As shown in Fig.~\ref{fig7}, for Contract $C$, Contract $B$ is the \emph{msg.sender} of this call, and User $A$ is the \emph{tx.origin} of this call.
    Using \emph{tx.origin} for authentication, malicious users can easily bypass authentication and steal ethers from the attacked contract after cheating your trust.
    The contract in Listing 5 is authenticated using \emph{tx.origin}. If the attacker induces the victim to transfer ethers to the attack contract shown in Listing 6 by various means, the attack contract can steal all deposits of the contract in Listing 5.
              \begin{lstlisting}[     language=C++,
                            breaklines=true,
                            captionpos=bc,
                            basicstyle = \footnotesize,
                            title={Listing 5: victim contract},
                            keywordstyle=\color{red},
                            commentstyle=\color{blue},
                            stepnumber=1
                            ]
contract Attacked{
    address public owner;
        
    constructor (address _owner){
        owner = _owner;
    }
        
    function withdrawAll(address _recipient) public{
        require(tx.origin == owner);
        _recipient.transfer(this.balance)
    }
}
    \end{lstlisting}
              \begin{lstlisting}[     language=C++,
                            breaklines=true,
                            captionpos=bc,
                            basicstyle = \footnotesize,
                            title={Listing 6: attack contract},
                            keywordstyle=\color{red},
                            commentstyle=\color{blue},
                            stepnumber=1
                            ]
import "Attack.sol";
    
contract Attacker{
    Attacked attacked;
    address attacker;
        
    constructor(Attacked _attacked, address _attacker){
        attacked = _attacked;
        attacker = _attacker;
    }
        
    function () public payable{
        attacked.withdrawAll(attacker);
    }
}
\end{lstlisting}

    
\noindent\emph{Missing constructor}. 
If the developer does not intend to write constructor function, the harm of this problem is very limited (eg., incomplete contract structure). But if the developer intends to write a constructor but write the wrong function name, then any user can call the function (eg., contract in Listing 7), which can cause serious security risks~\cite{fontein2018comparison,parizi2018empirical}. This paper recommends using \emph{constructor} keyword to declare constructors, which can effectively avoid the loss caused by the misspelling of constructor names. We check whether there is a constructor in each contract body. 

                  \begin{lstlisting}[     language=C++,
                            breaklines=true,
                            captionpos=bc,
                            basicstyle = \footnotesize,
                            title={Listing 7: harm of misspelling constructor name},
                            keywordstyle=\color{red},
                            commentstyle=\color{blue},
                            stepnumber=1
                            ]
 pragma solidity 0.5.0;
   
 contract Foo{
    address public owner;   //owner is the owner of the contract
    //Anyone cal now be the owner of the contract because the
    //function name is misspelled.
    function foo() public{
        owner = msg.sender;
    }
 }
    \end{lstlisting}

\noindent  \emph{Locked money}. If a contract needs to receive ethers, at least any function in the contract should be declared as \emph{payable}. At the same time, at least one statement should be included in the contract to enable the transfer ethers. Otherwise, all ethers in the contract account will be locked and can never be transferred. 


\noindent  \emph{Integer overflow}. The integer overflow problem exists widely in computer science. Because source based code analysis cannot accurately determine which statement may cause integer overflow, we adopt program instrumentation technique to prevent integer overflow through code insertion.

\noindent  \emph{Unsafe type inference}~\cite{tikhomirov2018smartcheck}. The keyword \emph{var} is provided in \emph{Solidity} language, which automatically assigns types to variables. In \emph{Solidity}, the type of variable is inferred to be the smallest type of storage space that can accommodate the initial value. As shown in Listing 8, the type of \textbf{i} is matched to \emph{uint8}, which is the type that can store an initial value of 0 and require the smallest storage space. Using \emph{var} as a variable matching type can have security risks. The loop shown in Listing 8 is an infinite loop, because \emph{uint8} can represent a maximum value of 255, and more than 255 will return to zero. In Ethereum, calls containing infinite loops are not packaged into blocks.

                          \begin{lstlisting}[     language=C++,
                            breaklines=true,
                            captionpos=bc,
                            basicstyle = \footnotesize,
                            title={Listing 8: infinite loop caused by unsafe type inference},
                            keywordstyle=\color{red},
                            commentstyle=\color{blue},
                            stepnumber=1
                            ]
            for (var i = 0; i <= 256; i++){
                //do something
            }
    \end{lstlisting}



\subsection{Performance Problems}
\noindent  \emph{byte\lbrack\ \rbrack}. \emph{byte\lbrack\ \rbrack} can play the role of the byte array, but this is a very wasteful storage space, which may lead to much gas consumption~\cite{SolidityDevelopDoc}. The recommendation approach is to use the \emph{bytes} type.

\noindent  \emph{Costly loop}. The user of calling a contract can specify the number of gases that this call carries before making the call. If gases are sufficient, the remaining gases will be returned by Ethereum after the call is completed. If gases are insufficient, the call will fail and Ethereum will not return the consumed gases. Loops that execute too many statements can lead to excessive costs for a call, and transactions with excessive costs may not be packaged into blocks, which means that transactions will never succeed.

\subsection{Hidden threats of coding problems}
\noindent \emph{Timestamp dependence}~\cite{jiang2018contractfuzzer}. Miners can control the mining time, thus gaining an unequal competitive advantage (eg., codes in Listing 9). Avoiding contract execution results depends on environmental variables. If necessary, environmental variables are costly to miners (eg., use \emph{block.difficulty}). 
Consequently, it is especially noteworthy that \emph{now} and \emph{block.timestamp} should not be used as parameters of cryptographic functions, so that the random number generated will be controlled by miners.
                      \begin{lstlisting}[     language=C++,
                            breaklines=true,
                            captionpos=bc,
                            basicstyle = \footnotesize,
                            title={Listing 9: guessing contract affected by miners},
                            keywordstyle=\color{red},
                            commentstyle=\color{blue},
                            stepnumber=1
                            ]
         if (now % 2 == 0)
           winner = addr1;
         else
           winner = addr2;
    \end{lstlisting}

\noindent  \emph{Token API violation}~\cite{tikhomirov2018smartcheck}. 
Ethereum allows the distribution of tokens.
Before May 7, 2019, Ethereum had more than 100,000 token contracts. 
The ERC20, ERC721, and ERC165~\cite{ERC-20TokenStandard,ERC-721TokenStandard,ERC-165TokenStandard} token standards are currently popular token standards that specify the most basic state variables, events, functions, and function return types in token contracts.
Throwing an exception in some functions that return a Boolean value is not recommended because throwing an exception prevents the caller from getting a return value, which can lead to dysfunction for the caller. When a function fails, it can tell execution failure by returning a Boolean value.

\noindent  \emph{Using fixed point number type}. \emph{Solidity} supports declaring variables of fixed point number type, but it cannot assign these variables or assign them to other variables~\cite{SolidityDevelopDoc},  so there is no need to use fixed point number type at all.

\noindent  \emph{Private modifier}~\cite{tikhomirov2018smartcheck}. The \emph{private} keyword is provided in Ethereum to indicate that the external visibility of a state variable or function is private. But unlike other programming languages, the use of \emph{private} does not make state variables and functions invisible to the outside world. Miners can view all the codes of the contract and the values of state variables, consequently the password in Listing 10 is available to the miner. 

                          \begin{lstlisting}[     language=C++,
                            breaklines=true,
                            captionpos=bc,
                            basicstyle = \footnotesize,
                            title={Listing 10: visible password },
                            keywordstyle=\color{red},
                            commentstyle=\color{blue},
                            stepnumber=1
                            ]
pragma solidity ^0.4.18;
    
contract Vault{
    bytes32 private password;
        
    function Vault(bytes32 _password) public payable{
        password = _password;
    }
        
    //Miners can take all the money of the contract
    function unlock(address _owner, bytes32 _password) public{
        if(password == _password){
            _owner.transfer(this.balance);
        }
    }
}
    \end{lstlisting}

\noindent \emph{Redundant refusal of payment}~\cite{tikhomirov2018smartcheck}. 
Starting from \emph{Solidity} 0.4.0, contracts without the fallback function will reject payment by default. This makes the function in Listing 11 redundant.

                          \begin{lstlisting}[     language=C++,
                            breaklines=true,
                            captionpos=bc,
                            basicstyle = \footnotesize,
                            title={Listing 11: redundant refusal of payment},
                            keywordstyle=\color{red},
                            commentstyle=\color{blue},
                            stepnumber=1
                            ]
    function() external payable{
        revert();
    }
    \end{lstlisting}

\noindent  \emph{Compiler version problem}~\cite{tikhomirov2018smartcheck}. The \textbf{$\wedge$} operator is provided in \emph{Solidity} to specify that this contract accepts compilation of the specified version number and its subsequent version compiler. However, the future development trend of \emph{Solidity} is unpredictable and may lead to semantic changes of some statements in future versions. In this way, we using the $\wedge$ symbol should be avoided
. Several methods for declaring the compiler version are shown in Listing 12. The recommended approach is to use the second or third.

                              \begin{lstlisting}[     language=C++,
                            breaklines=true,
                            captionpos=bc,
                            basicstyle = \footnotesize,
                            title={Listing 12: three ways to declare compiler versions},
                            keywordstyle=\color{red},
                            commentstyle=\color{blue},
                            stepnumber=1
                            ]
pragma solidity ^0.5.0; //bad: 0.5.0 and above
pragma solidity 0.5.0;  //good: only 0.5.0
pragma solidity >=0.5.0 <0.6.0; //best: 0.5.0 to 0.6.0
    \end{lstlisting}


\noindent  \emph{Style guide violation}~\cite{tikhomirov2018smartcheck}. In the official development document for \emph{Solidity}, the declaration and definition of functions, events, and arrays are standardized~\cite{SolidityDevelopDoc}. We think that the function name shown in the second line of code in Listing 13 is inappropriate because the two function names in Listing 13 do not allow one to understand the difference between their uses. It is recommended that function names begin with lowercase letters, event names begin with  uppercase letters, and there is no space between type and left brackets when array declarations are made.

                              \begin{lstlisting}[     language=C++,
                            breaklines=true,
                            captionpos=bc,
                            basicstyle = \footnotesize,
                            title={Listing 13: confusing function naming},
                            keywordstyle=\color{red},
                            commentstyle=\color{blue},
                            stepnumber=1
                            ]
    //the nameing of twn functions is confusing
    function transfer() public{ /*do something*/}
    function _transfer() public{ /*do something*/}
    \end{lstlisting}
    
\noindent\emph{Integer division}. The support for floating-point and decimal types in Ethereum is not perfect. All the results of integer division are rounded down, the use of integer division to calculate the number of ethers may cause economic losses, so it is try to avoid integer division.

\noindent  \emph{Implicit visibility level}~\cite{tikhomirov2018smartcheck}. Although \emph{Solidity} provides default visibility for each type of variable and function, explicitly specifying the visibility of each state variable and function improves the readability of the code.

\section{SolidityCheck}\label{Sec_Soliditycheck}

\subsection{Overview of \emph{SolidityCheck}}
    \begin{figure*}[t]
    \centering
    \includegraphics[scale=0.71]{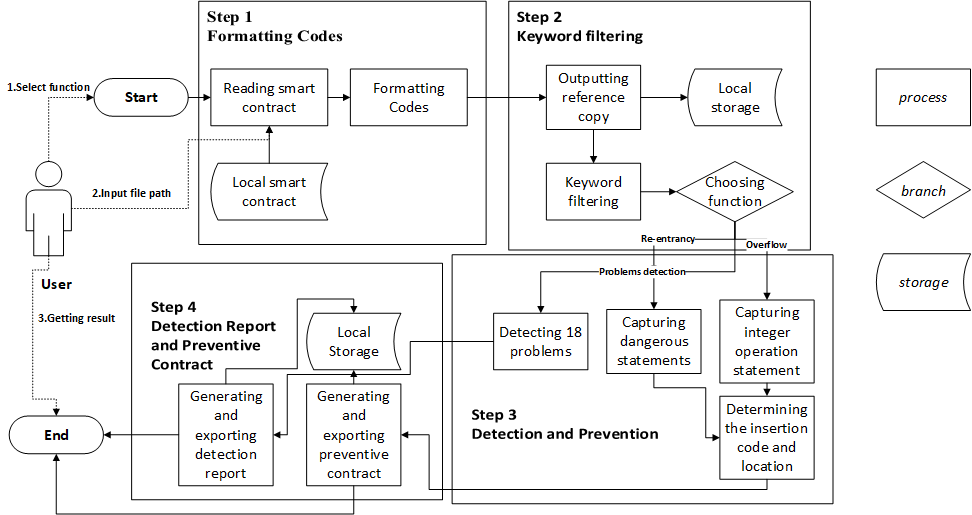}
    \caption{Overview of \emph{SolidityCheck}}
    \label{fig16}
    \end{figure*}
It is non-trivial to develop a tool that leverages regular expressions to locate the problems described in Section~\ref{sec_class} because of the following three reasons. 

First, the format of the source codes may not be suitable for the regular expressions, which usually handle code statements written in one line, and different programming habits make the source codes format different. The input source codes needs to be formatted to facilitate regular expression retrieval and matching. Consequently, we need implement the appropriate formatting method.

Second, regular matching of every line of codes will bring huge performance burden and make the retrieval efficiency extremely poor. Because regular expressions are usually based on the NFA (non-deterministic finite automaton) engine and are implemented by the ``matching backtracking" algorithm. However, NFA backtracking allows it to access the same state multiple times (if it arrives at that state through different paths). Therefore, in the worst case, it may be very slow to execute and spend a lot of CPU resources. To avoid the problem, we need to reduce the number of code statements that need regular matching without missing the problematic statements. We use \textbf{keyword filtering} to reduce unnecessary regular matching (statements with different kinds of problems always contain different specific characters).

Third, regular expressions are only suitable for detecting problems that exist within a single line of statements, but they are powerless for problems that span multiple lines of codes, such as \emph{costly loop}. Some programming tricks are needed to make the detection ability of regular expressions span multiple lines. For example, we use bracket matching to get the start and end positions of a loop statement. 

The main process of \emph{SolidityCheck} is divided into four steps, shown in Fig.~\ref{fig16}. The first step is \emph{formatting codes}, which enables regular expressions to easily detect the sentence characteristics of each code, and improves detection efficiency. 
The second step is \emph{keyword filtering}, \emph{SolidityCheck} extracts statements that may contain problems according to different keywords. Then, according to the functions selected by users, the filtered codes are processed in the third step, \emph{detection and prevention}. In this step, problems are detected or prevented according to the functions selected by users. The final step is \emph{detection report and preventive contract}, in this step problems detection report or preventive contract is output. The details of these steps are described in the following four subsections.

\subsection{Formatting Codes}\label{formatting_code}

The source code format of the smart contract is closely related to the programming habits of the developers, making the source codes in a variety of formats. The format in Listing 14 is Solidity's official recommendation for function header and its parameters declaration style. However, such kind of source codes is extremely unfriendly for regular expression matching because a complete statement has been written in several lines. Consequently, before retrieving the problematic statements, \emph{SolidityCheck} pre-processes the format of the source codes to write a sentence expressing complete semantics in one line. In general, the criteria for code formatting are described as follows:
\begin{enumerate}
\item All comments and blank lines in the original contract are filtered. The extra spaces in the statement are also eliminated.
 \emph{SoliditCheck} first stores the source codes in a string array by line, and then checks each item in the array sequentially. 
 If there is a "//" sub-string in one line, all comment characters in that line are replaced by "~" (including "//"); if there is a "/*" sub-string in one line, all characters are replaced by "~" before the next "*/" sub-string appears, and then "/*" and "*/" substrings are also be replaced by "~". After that, all comments in the source codes are filtered out, and then \emph{SolidityCheck} discards all blank lines and transfers the processed source codes to another string array.
 
\item Each formatted line of codes ends with a semicolon (;) or a left bracket (\{) or a right bracket (\}). 
In Solidity, the definitions of any contract header, function header and function modifier header are marked by left curly brackets (\{) to end the statement, the terminations of contract body, function body, and function modifier body are marked by right curly brackets (\}), while any other statement ends with semicolon (;).
\emph{SolidityCheck} gets the source codes into a string, replaces all line breaks with "~",. Then it scans the string sequentially, and adds a line break characters after encountering left brackets (\{), right brackets (\}) or semicolons (;). 
After this processing, all statements except for-statement are written in one line.

\item After the second step, a for-statement spans three lines, which is not conducive to regular expression matching. so the semicolon (;) in the for-statement is specially handled.  \emph{SolidityCheck} retrieves the for-statements, and then replaces the first two line breaks characters in the for statement with "~" so that the for-statement is written in one line.
\end{enumerate}

Listing 14 shows the codes before preprocessing, and Listing 15 shows the codes after the same contract is preprocessed. It is obvious that the number of lines of the codes to be detected is significantly reduced, and the format of the codes becomes easy for feature matching of regular expressions.
            \begin{lstlisting}[     language=C++,
                            breaklines=true,
                            captionpos=bc,
                            basicstyle = \footnotesize,
                            title={Listing 14: codes before contract preprocessing}
                            keywordstyle=\color{red},
                            commentstyle=\color{blue},
                            stepnumber=1
                            ]
    function deposit(
        address to,
        uint256 amount
        ){
        //Receiving address: to,Number: amount
        userBalance[to] += amount;
    }
    \end{lstlisting}
    
        \begin{lstlisting}[     language=C++,
                            captionpos=bc,
                            basicstyle = \footnotesize,
                            keywordstyle=\color{red},
                            commentstyle=\color{blue},
                            title={Listing 15: codes after contract preprocessing},
                            stepnumber=1
                            ]
    function deposit(address to,uint256 amount){
        userBalance[to] += amount;
    }
    \end{lstlisting}

\subsection{Keyword filtering}\label{keyword_filter}

\emph{Soliditycheck} does not regular match every line of codes, which greatly reduces the detection efficiency. We use \textbf{keyword filtering} to improve detection efficiency. 
There are specific sub-strings in regular expressions that we used to describe the characteristics of various problematic statements. 
A code statement that contains a certain type of string may have a type of problem. Through \textbf{keyword filtering}, we can effectively reduce the number of statements that need regular matching without increasing missed judgment. We illustrate \textbf{keyword filtering} with an example. The criteria for costly loop problems is described as follows:

\begin{itemize}
\item The conditional part of for-statement or while-statement contains function calls or identifiers.
\item for-statement or while-statement with a maximum number of statements executed exceeding 23 (The reason is described in Appendix B) .
\end{itemize}

From the above criteria, we can conclude that for-statements and while-statements may be costly loops and statements that do not include ``for" or ``while" must not be costly loops. Therefore, only the statements containing the keywords "for" and "while" can be matched with the features of costly loop problems.
\subsection{Detection and Prevention} \label{Sec_Detection}
This step includes two sub-steps: detection and prevention. For the detection step, we detect 18 kinds of problems besides re-entrancy vulnerability and integer overflow. For the prevention step, we use program instrumentation to prevent the re-entrancy vulnerability and integer overflow problem. The reason is that we cannot accurately detect these two problems through regular expression matching, but reporting all suspicious statements reduces the guidance of the detection results. Consequently, \emph{SolidityCheck} combines regular expressions and program instrumentation to achieve the purpose of prevention.
\subsubsection{Detection} \label{Sec_Detection}
During the detection step, \emph{SolidityCheck} matches problematic statements using regular expressions.
\emph{Soliditycheck} distributes formatted codes to 18 problem detection classes, each of which detects only one type of problems. In each problem detection class, the formatting codes are stored in an array of strings, and the program traverses line by line. The program first filters the code statements to be regularly matched by \textbf{keywords filtering}, and then matches the filtered statements according to the different regular expressions and corresponding detection rules we defined for each problem. If there is a problem, the line number of the codes is recorded, and the detection results of the 18 problem detection classes are summarized into a text file, which is the detect report.

\subsubsection{Prevention}
Prevention is further divided into re-entrancy vulnerability and integer overflow problems.\\
\emph{A. Re-entrancy vulnerability prevention.}\\
During the re-entrancy vulnerability prevention step, \emph{SolidityCheck} matches and inserts the codes using regular expressions.
Re-entrancy vulnerabilities are very special and dangerous. According to the harm of re-entrancy vulnerability, we divide re-entrancy vulnerability into the following two categories:
\begin{enumerate}
    \item Re-entrancy vulnerability with no ether transfer. This kind of re-entrancy vulnerabilities is called by a \emph{call} to the fallback function of the contract, carrying more than 2300 gas but not sending ether. Such vulnerabilities can cause the attack contract to repeatedly enter the attacked contract to perform operations, which may cause the state variable to be changed multiple times.
    \item Re-entrancy vulnerability with ether transfer. This kind of re-entrancy vulnerabilities are the most dangerous, usually resulting in a total loss of contract balance, and it has the following four characteristics:
    \begin{enumerate}
        \item Using \emph{call} to transfer ethers.
        \item Unrestricted gas.
        \item Deducing the balance after the transfer is completed
        \item \emph{Call} does not specify which function of the receiver will be called. Consequently, the contract uses the fallback function to respond to the transfer.
    \end{enumerate}
\end{enumerate}

Detecting a line or a piece of codes does not mean
accurately detecting re-entrancy vulnerabilities. Current smart contract detection tools cannot fully cover every re-entrancy vulnerability in every contract. Unfortunately, missing any re-entrancy vulnerability can cause a devastating blow to the contract. At present, the main way to detect re-entrancy vulnerabilities is to report every sentence that may introduce re-entrancy vulnerabilities, but this may lead to a lot of false positives and make the test results not instructive.


Our approach aims at the re-entrancy vulnerability with ether transfer. First, we define the \emph{dangerous statement}, which is the source of the re-entrancy vulnerability. A statement that contains the following three characteristics is called a \emph{dangerous statement}:
\begin{enumerate}
    \item Using \emph{call} to transfer ether.
    \item Unrestricted gas.
    \item \emph{Call} does not specify which function of the receiver will be called.
\end{enumerate}

\emph{Dangerous statements} are the source of re-entrancy vulnerabilities, but this does not mean that using \emph{dangerous statements}  necessarily includes re-entrancy vulnerabilities. Deducting the balance before the transfer can effectively avoid re-entrancy attacks. Codes in Listing 16 can prevent re-entrancy attacks.
 %
 %
 %
       \begin{lstlisting}[     language=C++,
          numbersep=1pt,
                            numbers=left,
                            breaklines=true,
                            captionpos=bc,
                            basicstyle=\footnotesize,
                            title={Listing 16: codes that effectively prevents re-entrancy}
                            keywordstyle=\color{red},
                            commentstyle=\color{blue},
                            stepnumber=1
                            ]
    function withdrawBalance_fixed(){
        //to protect against re-entrancy, the state variable
        //has to be changed before the call
        uint amount = userBalance[msg.sender];

        userBalance[msg.sender] = 0;    //First,deduction

        if(!(msg.sender.call.value(amount)())){
            throw;  //After,transfer
        }
    }
    \end{lstlisting}

We use regular expressions to define \emph{dangerous statements}: statements that match feature~\ref{4_1} or feature~\ref{4_2}.
    \makeatletter
    \@addtoreset{equation}{section}
    \makeatother
    \renewcommand{\theequation}{\arabic{section}.\arabic{equation}}
    \begin{equation}
    (.)+(\setminus .)(call)(\setminus .)(value)(\setminus ()(.)+(\setminus ))(\setminus ()(\setminus ))\label{4_1}
    \end{equation}
        \makeatletter
    \@addtoreset{equation}{section}
    \makeatother
    \renewcommand{\theequation}{\arabic{section}.\arabic{equation}}
    \begin{equation}
    (.)+(\setminus .)(call)(\setminus .)(value)(\setminus ()(.)+(\setminus ))(\setminus ()(\setminus '')(\setminus '')(\setminus ))\label{4_2}
    \end{equation}

Now we can match all the \emph{dangerous statements} with regular expressions. To prevent re-entrancy vulnerability, we need to insert specially constructed statements at some locations in the contract. The purpose of the insert statements is to terminate the operation before the transfer if there is transfer before deduction. 
The insertion statement will not interfere with the normal operation of the contract without the feature of first transfer and then deduction.

Second, for the convenience of describing the locations of the insertion statements, the concept of the function call chain is introduced. In Listing 17, function \emph{C} contains re-entrancy vulnerabilities, but it is an internal function and cannot be invoked by an external contract. This does not mean that the contract can be protected from re-entrancy attacks, because an attacker can invoke function  \emph{C} by calling function  \emph{A}, and can also launch re-entrancy attacks.
    \begin{lstlisting}[     language=C++,
                            breaklines=true,
                            captionpos=bc,
                            basicstyle=\footnotesize,
                            keywordstyle=\color{red},
                            commentstyle=\color{olive},
                            title={Listing 17: call chain schematic code},
                            stepnumber=1
                            ]
    contract example_1{
        mapping(address => uint256) userBalance;

        function A() public{
            B();
        }

        function B() internal{
            C();
        }

        function C() internal{
            msg.sender.call.value(1)();
            userBalance[msg.sender] -= 1;
        }
    }
    \end{lstlisting}
    Calls between functions constitute a call chain, and as long as any function in the chain has re-entrancy vulnerabilities, an attacker can achieve the effect of launching re-entrancy attacks by calling the prefix function in the chain of the function.

    In our approach, function  \emph{A} is called \emph{chain-head function}, and function  \emph{C} is called \emph{chain-tail function}. All \emph{chain-tail} functions are functions which contain \emph{dangerous statement}. \emph{Chain-tail functions} are called \emph{direct call function}, and other functions in the chain are called \emph{indirect call function} except \emph{chain-tail functions}.
    
    By defining the \emph{call chain}, we can accurately describe where the prevention statements (we call the prevention statements \emph{vaccines}) are inserted. 
    
    Then, we define the \emph{ledger}: the variable used to record the correspondence between the address and the number of tokens the address holds is called the \emph{ledger}. There are many variables in a contract that can act as a \emph{ledger}, and due to the lack of available tools to determine which variable is the \emph{ledger}. Therefore, we set the \emph{ledger} as the mapping (address $=>$ uint256) variable of the first declaration in the contract, which is based on our own experience.
    
    Now, we describe the structure of 4 \emph{vaccines} and the insertion positions of different \emph{vaccines} in table~\ref{Tab_Insertion}. The original codes move backwards after insertion.

\begin{table}[h]
\caption{Insertion code composition and insertion location}
\label{Tab_Insertion}
\begin{tabular*}{9cm}{p{1.3cm}<{\centering}|p{4cm}<{\centering}p{2.3cm}}
\toprule
Code type & Composition structure & Insertion position \\
\midrule
A & if(Bexe == 0) \{ Bexe = ledger[etherReceiver];\} & first line of direct call functions and first line of chain-head functions \\
\midrule
B & Aexe = ledger[etherReceiver]; require(Aexe\textless Bexe); & The front line of a dangerous statement of direct call functions \\
\midrule
C & Aexe = 0; Bexe = 0; & The next line of a dangerous statement of direct call functions \\
\midrule
D & uint256 Aexe = 0; uint256 Bexe = 0; & First line of the contract \\
\bottomrule
\end{tabular*}
\end{table}

Table~\ref{Tab_Insertion1} explains the purpose of inserting four types of code.
\begin{table}[h]
\centering
\caption{Insertion purpose of 4 kinds of codes}
\label{Tab_Insertion1}
\begin{tabular*}{8cm}{p{1.3cm}<{\centering}|p{6cm}}
\toprule
Code type & Insertion purpose  \\
\midrule
A & Number of tokens held at the ethers receiving address at the beginning of the transfer business  \\
\midrule
B & Using to obtain the number of tokens held at the address of receiving ethers before the transfer is initiated. If the number of tokens does not decrease at this time, execution will be aborted \\
\midrule
C & Resetting Aexe, Bexe   \\
\midrule
D & Declaring Aexe, Bexe \\
\bottomrule
\end{tabular*}
\end{table}

By inserting four \emph{vaccines}, \emph{vaccines} are able to abort and roll back the operation on time if the contract used dangerous statements and contained the feature \textbf{balance deducted after transfer}.

Because the re-entrancy vulnerability prevention function does not apply to all contracts (e.g., those contracts that do not declare \emph{ledger}), to facilitate use, we list the switch of this function separately. In this way, developers can use it according to their circumstances.

Besides, \emph{SolidityCheck} inserts functions (named deposit\_test) into the contract. Calling the function in a private chain environment and sending enough ethers can achieve the effect of detecting re-entrancy vulnerabilities (by observing the results of function execution).

\emph{B. Integer overflow problem prevention}.\\
During the integer overflow problem prevention step, \emph{SolidityCheck} matches and inserts the codes using regular expressions.

The BEC project was officially launched on February 23, 2018, with a maximum market value of more than \$28 billion. However, two months after its launch, the attacker found that there was an integer overflow problem in the BEC contract, and launched an attack against the problem, leading to an unlimited issue of BEC tokens, which eventually triggered a wave of selling. The final result was that the market value of BEC token was almost zero.

Listing 18 shows the BEC source codes~\cite{BecToken} with an integer overflow problem, and in line 3 there is an integer overflow problem.

       \begin{lstlisting}[     language=C++,
                            breaklines=true,
                            captionpos=bc,
                            basicstyle=\footnotesize,
                            keywordstyle=\color{red},
                           numbers = left,
                                     numbersep=1pt,
                           commentstyle=\color{blue},
                            title={Listing 18: BEC source code},
                            stepnumber=1
                            ]
 function batchTransfer(address[] _receivers, uint256 _value) public whenNotPaused returns (bool) {
    uint cnt = _receivers.length;
    uint256 amount = uint256(cnt) * _value;
    require(cnt > 0 && cnt <= 20);
    require(_value > 0 && balances[msg.sender] >= amount);

    balances[msg.sender] = balances[msg.sender].sub(amount);
    for (uint i = 0; i < cnt; i++) {
        balances[_receivers[i]] = balances[_receivers[i]].add(_value);
        Transfer(msg.sender, _receivers[i], _value);
    }
    return true;
}
    \end{lstlisting}

Detection of integer overflow depends on logic analysis and semantics understanding of codes. It is difficult to determine that any integer operation statement has the risk of integer overflow.

At present, the common method to prevent integer overflow in the Ethereum is to use \emph{SafeMath} library~\footnote{https://ethereumdev.io/safemath-protect-overflows/} for integer operation. In Listing 18, lines 7 to 9 use \emph{SafeMath} library functions for addition and subtraction. The \emph{SafeMath} library has several versions of implementations, and part codes of the most popular implementation are shown in Listing 19~\cite{SafeMath}.

       \begin{lstlisting}[     language=C++,
                            breaklines=true,
                            captionpos=bc,
                            basicstyle=\footnotesize,
                            keywordstyle=\color{red},
                           numbers = left,
                                     numbersep=1pt,
                           commentstyle=\color{blue},
                            title={Listing 19: the part codes of SafeMath},
                            stepnumber=1
                            ]
 function add(uint256 a, uint256 b) internal pure returns (uint256) {
    uint256 c = a + b;
    require(c >= a, "SafeMath: addition overflow");

    return c;
    }
    \end{lstlisting}
The integer overflow prevention draws on the idea of \emph{SafeMath} library. For each integer operation statement, different verification codes are inserted before and after the statement to verify whether the results of this integer operation is correct, and the verification code can terminate in time after the overflow occurs. It is equivalent to actively implement the function of \emph{SafeMath} library. When the program is implemented, the program will not add the verification codes in the \emph{SafeMash} library to avoid unnecessary gas consumption.

First, \emph{SolidityCheck} captures statements with the characteristics of~\ref{4_3} or~\ref{4_4}:
    \makeatletter
    \@addtoreset{equation}{section}
    \makeatother
    \renewcommand{\theequation}{\arabic{section}.\arabic{equation}}
    \begin{equation}
    \begin{split}
    &\wedge (\setminus s)*(\setminus w)*(\setminus s)+((\setminus w)|(\setminus ()|(\setminus ))|(\setminus [)|(\setminus ])|(\setminus .))\\ &+(\setminus s)*(\setminus =)(\setminus s)*((\setminus w)|(\setminus ()|(\setminus ))|(\setminus [)|(\setminus ])|(\setminus .))+\\ &(\setminus s)*((\setminus +)|(\setminus -)|(\setminus *)|(\setminus /)|(\setminus \%))(\setminus s)*((\setminus w)|(\setminus ()|\\ &(\setminus ))|(\setminus [)|(\setminus ])|(\setminus .))+(\setminus s)*(;)\$\label{4_3}
    \end{split}
    \end{equation}
    \makeatletter
    \@addtoreset{equation}{section}
    \makeatother
    \renewcommand{\theequation}{\arabic{section}.\arabic{equation}}
    \begin{equation}
    \begin{split}
    &\wedge (\setminus s)*(\setminus w)*(\setminus s)+((\setminus w)|(\setminus ()|(\setminus ))|(\setminus [)|(\setminus ])|(\setminus .))\\ &+(\setminus s)*(((\setminus +)(\setminus =))|((\setminus -)(\setminus =))|((\setminus *)(\setminus =))|((\setminus \\ & /)(\setminus =))|((\setminus \%)(\setminus =)))(\setminus s)*((\setminus w)|(\setminus ()|(\setminus ))|(\setminus [)|(\setminus ]\\ &)|(\setminus .))+(\setminus s)*(;)\$\label{4_4}
    \end{split}
    \end{equation}
    The sentence structure defined in Formula~\ref{4_3} is as follows:
    \makeatletter
    \@addtoreset{equation}{section}
    \makeatother
    \renewcommand{\theequation}{\arabic{section}.\arabic{equation}}
    \begin{equation}
    ope1 = ope2 (+|-|*|/|\%) ope3\label{4_5}
    \end{equation}
    The sentence structure defined in Formula~\ref{4_4} is as follows:
     \makeatletter
    \@addtoreset{equation}{section}
    \makeatother
    \renewcommand{\theequation}{\arabic{section}.\arabic{equation}}
    \begin{equation}
    ope1 ((+=)|(-=)|(*=)|(/=)|(\%=)) ope2\label{4_6}
    \end{equation}

For statements with different operations, the composition of the preventive codes inserted after the statement is shown in Table~\ref{Tab_Relation}.
\begin{table}
\centering
\caption{The Corresponding Relation between Integer Operational Code and Preventive Code(after)}
\label{Tab_Relation}
\begin{tabular*}{9cm}{p{3cm}<{\centering}p{5cm}<{\centering}}
\toprule
\textbf{Statement type} & \textbf{Preventive code (after) composition architecture} \\
\midrule
$ope1 = ope2 + ope3;$ & $require(ope1 > = ope2);$ \\
\midrule
$ope1 = ope2 - ope3;$ & $require(ope1 < = ope2);$ \\
\midrule
$ope1 = ope2 * ope3;$ & $require(ope2 == 0 || ope1 / ope2 == ope3);$ \\
\midrule
$ope1 = ope2 / ope3; $& $require(ope3>0); $\\
\midrule
$ope1 = ope2 \% ope3;$ & $require(ope3 != 0);$ \\
\midrule
$ope1 += ope2; $& $require(ope1 > = ope2);$ \\
\midrule
$ope1 -= ope2; $& $require(anti\_overflow\_temp\_count >= ope2); $\\
\midrule
$ope1 *= ope2;$ & $require(ope2 == 0 || ope1 / ope2 == anti\_overflow\_temp\_count);$ \\
\midrule
$ope1 /= ope2;$ &$ require(ope2 > 0);$ \\
\midrule
$ope1 \%= ope2;$ & $require(ope2 != 0);$ \\
\bottomrule
\end{tabular*}
\end{table}

Table~\ref{before} shows the two statement types and the code inserted before the statement. Among them, the \textbf{count} part of variable naming is an integer that grows by 1 for each insertion of code into a particular statement, starting from 1, to prevent repeated declarations of variables. If the code is inserted before and after the integer operation code, the variable names before and after the code correspond.

As with the function of preventing re-entrancy vulnerability, it is not appropriate to insert codes into each tested contract (increasing gas consumption). Consequently, we also list the functions of preventing integer overflow separately, and users can choose according to their situation.

\begin{table}
\centering
\caption{The Corresponding Relation between Integer Operational Code and Preventive Code(before)}
\label{before}
\begin{tabular*}{9cm}{p{3cm}<{\centering}p{5cm}<{\centering}}
\toprule
\textbf{Statement type} & \textbf{Preventive code (before) composition architecture} \\
\midrule
$ope1 *= ope2;$ & $uint256\ anti\_overflow\_temp\_count = ope1;$ \\
\midrule
$ope1-= ope2;$ & $uint256\ anti\_overflow\_temp\_count = ope1;$ \\
\bottomrule
\end{tabular*}
\end{table}

\subsection{Detection Report and Preventive Contract}

\emph{SolidityCheck} outputs two different files depending on the functionality selected by the user.
\begin{enumerate}
\item \emph{Detection report}. That is the detection report of 18 kinds of problems besides re-entrancy vulnerability and integer overflow problem.
\item \emph{Preventive contracts}. A contract that prevents problems after inserting code.
\end{enumerate}
A preventive contract is a contract after inserting codes base on the original contract. \emph{SolidityCheck} prints out the number of lines of inserted codes to help users to know the location.

Listing 20 shows the specific format of the test report.
    \begin{lstlisting}[     language=XML,
                            breaklines=true,
                            title={Listing 20: detection report format},
                            captionpos=bc,
                            basicstyle=\footnotesize,
                            keywordstyle=\color{blue},
                            commentstyle=\color{blue}
                            ]
<Detect Report>
    <Reporting information>
        <Smart contract file path/>
        <Number of lines of original contract code/>
        <Detection time/>
        <Total number of problematic  statements/>
    </Reporting information>
    <Details of the  problem>
        < Problem number/>
        < Problem name/>
        < Problem code line number/>
         < Problem description/>
        <Suggested modifications/>
    </Details of the  problem>
    <!-- Eliminate the details of the next 17  problems -->
</Detect Report>
\end{lstlisting}

Listing 21 shows a prevention contract (the original contract: \emph{Reentrancy.sol} from \emph{not-so-smart-contracts}~\cite{NotSoSmartContracts}). As shown in listing 21, \emph{SolidityCheck} does not insert codes in function \emph{withdrawBalanc\_fixed2}, the inserted codes prevent the re-entrancy vulnerability in function \emph{withdrawBalance} and the inserted codes do not affect the execution of function \emph{withdrawBalanc\_fixed}, while the inserted function \emph{deposit\_test} detects the re-entrany vulnerability.

    \begin{lstlisting}[     language=C++,
                            breaklines=true,
                            title={Listing 21: A contract to prevent re-entrancy vulnerability},
                            captionpos=bc,
                            basicstyle=\footnotesize,
                            keywordstyle=\color{blue},
                            commentstyle=\color{blue}
                            ]
 pragma solidity ^0.4.15;
 contract Reentrance {
    uint256 public Aexe=0;
    uint256 public Bexe=0;
    mapping (address => uint) userBalance;
    function getBalance(address u) constant     returns(uint){
        return userBalance[u];
    }
    function addToBalance() payable{
        userBalance[msg.sender] += msg.value;
    }
    function withdrawBalance(){
	if(Bexe==0){
	    Bexe=userBalance[msg.sender];
	}
	Aexe=userBalance[msg.sender];
	require(Aexe<Bexe);
        if(!(msg.sender.call.value(userBalance[msg.sender])())){
	   Aexe=0;
	   Bexe=0;
        throw;
        }
        userBalance[msg.sender] = 0;
    }
    function withdrawBalance_fixed(){
        if(Bexe==0){
	    Bexe=userBalance[msg.sender];
	}
        uint amount = userBalance[msg.sender];
        userBalance[msg.sender] = 0;
	Aexe=userBalance[msg.sender];
	require(Aexe<Bexe);
        if(!(msg.sender.call.value(amount)())){
	    Aexe=0;
	    Bexe=0;
            throw;
        }
    }
    function withdrawBalance_fixed_2(){
    msg.sender.transfer(userBalance[msg.sender]);
        userBalance[msg.sender] = 0;
    }
    function deposit_test() public payable{
	userBalance[msg.sender]+=msg.value;
	withdrawBalance();
	withdrawBalance_fixed();
    }
}

\end{lstlisting}

\section{Experimental Design}\label{Sec_Experimental}
\subsection{Research Questions}\label{Sec_Experimental_questions}
In this section, a series of experiments is conducted to validate \emph{SolidityCheck} based on a large amount of data sets collected by us.
The purpose of the experiments is to explore the following four research questions:
\begin{itemize}
    \item  RQ1: Can \emph{SolidityCheck} detect any smart contract written in \emph{Solidity} language and correctly output detection reports? 
\item RQ2: Is \emph{SolidityCheck} more efficient than similar tools?
\item RQ3: Is the detection quality of \emph{SolidityCheck} better than that of similar tools? 
\item RQ4: Can \emph{SolidityCheck} prevents important vulnerabilities such as re-entrancy vulnerabilities and integer overflow problems?
\end{itemize}
We designed RQ1 to validate the usability of \emph{SolidityCheck}. RQ2 is used to investigate whether the detection efficiency of \emph{SolidityCheck} is higher than that of similar tools. RQ3 is used to verify the detection quality of \emph{SolidityCheck}. We use recall and precision to judge the detection quality. RQ4 is used to validate the effectiveness of our problem prevention ability.
\subsection{Experimental Data Set}
Using Web crawler technology, we collected 1363 smart contracts written in \emph{Solidity} from the \textbf{etherscan.io}\cite{EtehreumBlockChainExplorer}, totaling 1,239,927 lines of codes. To understand the size of the contract in the data set, we counted the number of code lines of each contract, and the results are shown in Table~\ref{codelines}.

\begin{table}[h]
\centering
\caption{Distribution of contract code lines}
\label{codelines}
\begin{tabular}{c|cc}
\toprule
 Size & Number & Proportion\\
 \midrule
 0-500 & 699 & 51.3\% \\
 \midrule
 500-1000 & 334 & 24.5\%  \\
 \midrule
 1000+ & 329 & 24.2\% \\
\bottomrule
\end{tabular}
\end{table}

\subsection{Contrast Tools}

Based on the theory described in Section~\ref{Sec_Soliditycheck}, we implemented a tool named \emph{SolidityCheck} for Solidity language, which is now open source~\footnote{https://github.com/xf97/SolidityCheck}. 
Appendix C provides the implementation details of the tool.

To measure \emph{SolidityCheck}'s capability, we select several state-of-the-art tools for comparison. Because the software and security engineering research community has relied on free and open source software~\cite{parizi2018empirical} for a long time, we choose comparison tools from open source projects on github~\cite{Github}. According to their popularity, We select the following eight tools:
\begin{enumerate}
    \item\emph{Remix}\cite{Remix}. The \emph{Solidity}integrated development environment officially recommended by Ethereum. We use the \emph{Solidity static analysis} function of \emph{Remix}.
    \item \emph{Mythx}~\cite{Mythril}. Security analysis tool for EVM bytecode. Supporting smart contracts built for Ethereum. \emph{MythX} uses \emph{Mythril} and other non open source tools like a static analysis tool - \emph{Maru} and a greybox fuzzer - \emph{Harvey}, so it detects a wide range of vulnerabilities. There are many implementations available for \emph{Mythx}, and we use \emph{Pythx} (a library for the \emph{MythX} smart contract security analysis platform.).
    \item \emph{Oyente}\cite{Oyente}. The commonly used static code analysis tool for Ethereum smart contracts was put forward earlier in the same kind of tools~\cite{luu2016making}.
    \item \emph{Solhint}~\cite{Solhint}. \emph{Solhint} is an open source project for linting \emph{Solidity} code. This project provides both security and style Guide validations.
    \item \emph{Securify}~\cite{Securify}. The project sponsored by the Ethereum Foundation is a security scanner for Ethereum smart contracts~\cite{tsankov2018securify}.
    \item \emph{SmartCheck}~\cite{smartCheck}, Static code analysis tool for Ethereum smart contracts~\cite{tikhomirov2018smartcheck}.
    \item \emph{ContractFuzzer}~\cite{ContractFuzzer}. The Ethereum smart contract fuzzer for security vulnerability detection~\cite{jiang2018contractfuzzer}.
    \item \emph{Osiris}~\cite{Osiris}. A tool to detect integer bugs in Ethereum smart contracts~\cite{torres2018osiris}.
\end{enumerate}

\subsection{Experimental Environment}
Our experimental environment is a computer running Ubuntu (Ubuntu 16.04) system. The memory is 32GB, the CPU is Inter Xeon Silver 4108, and the GPU is NVIDIA Quadro P4000. In the quality experiment, we installed different versions of \emph{solc} through \emph{docker} and obtained the bytecode file of each contract.

\subsection{Experimental Results}
\subsubsection{Usability}
To answer RQ1, we used the problem detection function of \emph{SolidityCheck} to test 1363 smart contracts and verify whether \emph{SolidityCheck} correctly outputs the detection report of each smart contract. The experimental results show that \emph{SolidityCheck} correctly outputs the detection report of each smart contract. Furthermore, we use the following indicators to measure the proportion of problems in the experimental data set. 
\emph{ContractsNumber$_{this\_problem}$} is the number of contracts containing one problem, and \emph{NumberOfTestContracts} is the number of contracts being tested.
    \makeatletter
    \@addtoreset{equation}{section}
    \makeatother
    \renewcommand{\theequation}{\arabic{section}.\arabic{equation}}
    \begin{equation}
    \begin{split}
    Proportion = &ContractsNumber_{this\_problem}/ \\ &NumberOfTestContracts
    \end{split}
    \end{equation}

\begin{figure}[h]
    \centering
    \includegraphics[scale=0.60]{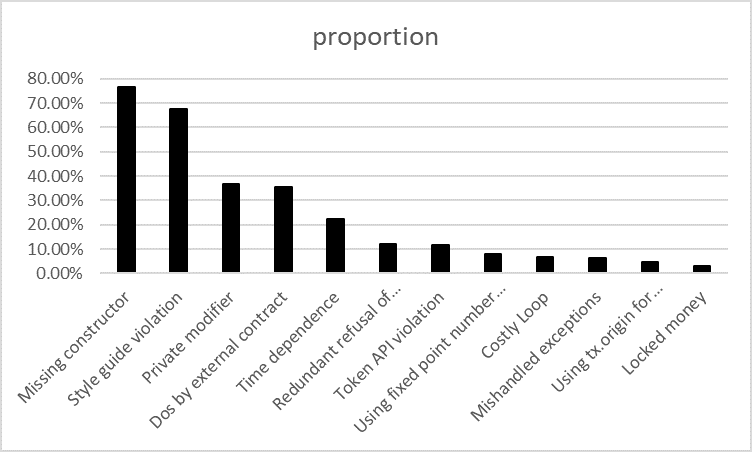}
    \caption{The proportion of 12 problems in experimental data sets}
    \label{fig23}
    \end{figure}

According to our experiments, most of the contracts have problems, 61\% of them have reported 10 or more problems. Because the number of different problems varies greatly in the data set, we present some maximal or minimal values as follows: \emph{Compiler version problem}--99.8\%, \emph{Implicit visibility level}--82.1\%, 
\emph{Using fixed point number type}--0.1\%, \emph{byte\lbrack\ \rbrack}--0\%, \emph{Balance equality}--0\%, \emph{Unsafe type infefrence}--0\%. The proportion of the remaining 11 problems in the data set is shown in Fig.~\ref{fig23}.
Since most contracts do not introduce a new version of the security specification statement into the contract, most contracts are reported to contain this problem. The number of the hidden threats of coding problems is also high. Compared with security problems, the external visibility of state variables and functions is trivial and less valued, consequently, it is common in smart contracts. The last few problems are about statements that are used less frequently, or that most people know are wrong, and the number of occurrences is very small. This is also in line with our expectations before the test, and to some extent reflects the credibility of \emph{SolidityCheck}.

More importantly, \emph{SolidityCheck} found two contracts with re-entrancy vulnerabilities, 18 contracts with miner-controlled random numbers and 10 contracts with locked money. After our manual verification, all problems are true and have not been reported before.
\subsubsection{Efficiency}\label{exp_efficiency}
To answer RQ2, we design a series of experiments to record the time consumed and also compare with other tools. For bytecode-based tools such as \emph{Oyente}, the measure of detection efficiency is the average detection time of each contract. For source-based tools such as \emph{SmartCheck}, the measure of detection efficiency is the number of lines of codes detected per second. In this way, we design two groups of experiments. 
In the first group, we used all the tools to test the 50 contracts we randomly selected to measure the average time each tool spent per contract. In the second group, we used \emph{SolidityCheck} and \emph{SmartCheck} to test the whole experimental data set and measure the number of lines of codes per second for each tool. 

\textbf{Experiment 1}. Because some of the tools have exceptions when contracts are analyzed, and these exceptions can cause analysis interruptions. Consequently, we use the quotient of the total time and the number of contracts successful analyzed as the average time. Table~\ref{Tab_ave_time} shows the average detection time for each tool to detect a contract.~\footnote{\emph{Remix}: we count the run time of compilation and solidity static analysis}~\footnote{\emph{Mythx}: we use \emph{Pythx} and only count runTime}

\begin{table*}[t]
\centering
\caption{Average detecting time for a contract}
\label{Tab_ave_time}
\begin{tabular}{c|ccccccccc}
\toprule
 & \emph{ContractFuzzer} & \emph{Mythx} & \emph{Osiris} & \emph{Oyente}  & \emph{Securify} &  \emph{Remix}  & \emph{SmartCheck} & \emph{Solhint} & \emph{SolidityCheck}    \\
\midrule
\textbf{time(s)} & 135.5 &  86.7  & 80.90  & 56.27 & 52.14  & 7.63 & 1.83  & 0.57 &  0.23  \\
\bottomrule
\end{tabular}
\end{table*}

\begin{figure}[h]
    \centering
    \includegraphics[scale=0.65]{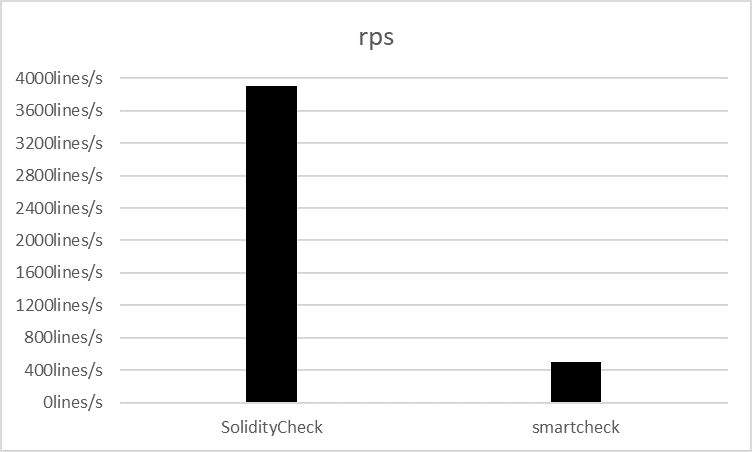}
    \caption{Performance performance of \emph{SmartCheck} and \emph{SolidityCheck}}
    \label{fig24}
    \end{figure}

\textbf{Experiment 2}. We define the following indicator to measure the detection efficiency of different tools. The \emph{rps} value is a quotient of the total number of lines of codes for all tested contracts and the total time consumed to test all contracts.
    \makeatletter
    \@addtoreset{equation}{section}
    \makeatother
    \renewcommand{\theequation}{\arabic{section}.\arabic{equation}}
    \begin{equation}
    \begin{split}
    rps = &\sum{NumberOfLines_{eachContract}} / \\ & \sum{timeConsuming_{eachContract}}
    \end{split}
    \end{equation}
The experimental result shows that \emph{SmartCheck} takes 2491.33 seconds to test the whole experimental data set, while \emph{SolidityCheck} takes 318.02 seconds. The \emph{rps} values of the two tools are shown in Fig.~\ref{fig24}. The detection efficiency of \emph{SolidityCheck} is 683.4\% higher than that of \emph{SmartCheck}.
\subsubsection{Quality}
To answer RQ3, we design a series of experiments to measure \emph{recall} and \emph{precision} of \emph{SolidityCheck}. We randomly selected 10 contracts from the entire experimental data set as test cases to accurately measure the quality of each tool. We determine the number of actual problems in each contract by manual review. To ensure the accuracy of manual review, we invite trained researchers to assist in identifying problems in the contract.

First, several experimental indicators are defined, as shown in Table~\ref{Tab_Definitions}.
These indicators refer to:
\begin{itemize}
    \item TP (True Positive) means the problem actually exists and the tool report exists.
    \item FN (Fasle Negative) means the problem does not exist but the tool report exists.
    \item FP (False Positive) means the problem actually exists but the tool report does not exist.
\end{itemize}
For Ethereum smart contracts, missed judgment is much more serious than misjudgment. Because smart contracts are mostly hundreds of lines of codes, it is not difficult to manually review each reported problem, but missing any one of them could be a fatal blow to smart contracts.
\begin{table}[h]
\centering
\caption{Definitions of accurate detection, misjudgement and misjudgement}
\label{Tab_Definitions}
\begin{tabular*}{9cm}{p{2.5cm}<{\centering}|p{3cm}<{\centering}|p{2.5cm}<{\centering}}
\toprule
 & \textbf{Actual existence} & \textbf{Non-existence} \\
\midrule
\textbf{Detect problem} & TP & FN \\
\midrule
\textbf{Not detected} & FP & TN \\
\bottomrule
\end{tabular*}
\end{table}

Formulas~\ref{recall_rate} and~\ref{precision_rate} give the definition of \emph{recall} and \emph{precision} of a single contract.
    \makeatletter
    \@addtoreset{equation}{section}
    \makeatother
    \renewcommand{\theequation}{\arabic{section}.\arabic{equation}}
    \begin{equation}
    Recall\;rate = (TP) / (TP+FP)\label{recall_rate}
    \end{equation}
        \makeatletter
    \@addtoreset{equation}{section}
    \makeatother
    \renewcommand{\theequation}{\arabic{section}.\arabic{equation}}
    \begin{equation}
    Precision\;rate = (TP) / (TP+FN)\label{precision_rate}
    \end{equation}
    
The \emph{recall} and \emph{precision} for each tool are presented in Table~\ref{Tab_results}. In Table~\ref{Tab_results}, the \emph{error} means an error occurred when the tool detected the contract so that it could not output the analysis result, and the \emph{N/A} represents the numerator is zero. 
    
\begin{table*}[t]
\centering
\caption{ Comparative results of recall and precision for different tools}
\label{Tab_results}
\begin{tabular}{| p{0.8cm}<{\centering} |p{1.6cm}<{\centering}|p{1.8cm}<{\centering}|p{1cm}<{\centering}|p{1cm}<{\centering}|p{1cm}<{\centering}|p{1.1cm}<{\centering}|p{1.2cm}<{\centering}|p{1.3cm}<{\centering}|p{1cm}<{\centering}|p{1.5cm}<{\centering}|}
\toprule
 Test case &  & \emph{ContractFuzzer} & \emph{Mythx} & \emph{Osiris} & \emph{Oyente}  & \emph{Securify} &  \emph{Remix}  & \emph{SmartCheck} & \emph{Solhint} & \emph{SolidityCheck}    \\
\midrule
\multirow{3}*{1}& TP/FP/FN & 0/37/0 &  1/36/1  & error  & error & 0/37/4  & 19/18/2 & 21/16/0  & 0/37/0 &  37/0/1  \\
~ & recall(\%) & 0 & 2.7 & N/A & N/A & 0 & 51.4 & 56.8 & 0 & 100 \\
~ & precision(\%) & N/A & 50 & N/A & N/A & 0 & 90.47 & 100 & N/A & 97.4  \\
\midrule
\multirow{3}*{2}& TP/FP/FN & 0/4/0 &  1/3/2  & error  & error & 0/4/1  & 0/4/2 & 4/0/0  & 0/4/0 &  4/0/4  \\
~ & recall(\%) & 0 & 25 & N/A & N/A & 0 & 0 & 100 & 0 & 100 \\
~ & precision(\%) & N/A & 33.3 & N/A & N/A & 0 & 0 & 100 & N/A & 50  \\
\midrule
\multirow{3}*{3}& TP/FP/FN & 0/5/0 &  0/5/10  & error  & error & 0/5/1  & 0/5/0 & 1/4/1  & 0/5/0 &  3/2/0  \\
~ & recall(\%) & 0 & 0 & N/A & N/A & 0 & 0 & 20 & 0 & 60 \\
~ & precision(\%) & N/A & N/A & N/A & N/A & 0 & N/A & 100 & N/A & 97.4  \\
\midrule
\multirow{3}*{4}& TP/FP/FN & 0/79/0 &  0/79/1  & error  & error & 0/79/2  & 1/78/1 & 42/37/4  & 0/79/0 &  79/0/8  \\
~ & recall(\%) & 0 & 0 & N/A & N/A & 0 & 1.3 & 53.2 & 0 & 100 \\
~ & precision(\%) & N/A & 0 & N/A & N/A & 0 & 50 & 91.3 & N/A & 90.8  \\
\midrule
\multirow{3}*{5}& TP/FP/FN & 0/28/0 &  1/27/0  & 0/28/0   & 0/28/3 & 0/28/0  & 0/28/0 & 19/9/0  & 0/28/0 &  27/1/1  \\
~ & recall(\%) & 0 & 3.6 & 0 & 0 & 0 & 0 & 67.9 & 0 & 96.4 \\
~ & precision(\%) & N/A & 100 & N/A & 0 & N/A & 0 & 100 & N/A & 96.4  \\
\midrule
\multirow{3}*{6}& TP/FP/FN & 0/4/0 &  1/3/1  & error & error & error  & 2/2/2 & 2/2/2  & 0/4/0 &  4/0/8  \\
~ & recall(\%) & 0 & 25 & N/A & N/A & N/A & 50 & 50 & 0 & 100 \\
~ & precision(\%) & N/A & 50 & N/A & N/A & N/A & 50 & 50 & N/A & 33.3  \\
\midrule
\multirow{3}*{7}& TP/FP/FN & 0/8/0 &  1/7/2  & error  & error & 0/8/2  & 1/7/1 & 4/4/0  & 0/8/0 &  7/1/2  \\
~ & recall(\%) & 0 & 12.5 & N/A & N/A & 0 & 12.5 & 50 & 0 & 87.5 \\
~ & precision(\%) & N/A & 33.3 & N/A & N/A & 0 & 50 & 100 & N/A & 77.8  \\
\midrule
\multirow{3}*{8}& TP/FP/FN & 1/19/0 &  3/17/0  & error  & error & 0/20/1  & 0/20/2 & 6/14/0  & 0/20/0 &  17/3/1  \\
~ & recall(\%) & 5 & 15 & N/A & N/A & 0 & 0 & 30 & 0 & 85 \\
~ & precision(\%) & 100 & 100 & N/A & N/A & 0 & 0 & 100 & N/A & 94.4 \\
\midrule
\multirow{3}*{9}& TP/FP/FN & 0/45/0 &  0/45/0  & error  & error & 0/45/1  & 0/45/0 & 18/27/2  & 0/45/0 &  45/0/0  \\
~ & recall(\%) & 0 & 0 & N/A & N/A & 0 & 0 & 40 & 0 & 100 \\
~ & precision(\%) & N/A & N/A & N/A & N/A & 0 & N/A & 90 & N/A & 100  \\
\midrule
\multirow{3}*{10}& TP/FP/FN & 0/2/0 &  1/1/0  & error  & error & 0/2/1  & 0/2/0 & 1/1/0  & 0/2/0 &  2/0/2  \\
~ & recall(\%) & 0 & 50 & N/A & N/A & 0 & 0 & 50 & 0 & 100 \\
~ & precision(\%) & N/A & 100 & N/A & N/A & 0 & N/A & 100 & N/A & 50  \\
\midrule
\multirow{3}*{Overall}& TP/FP/FN & 1/231/0 &  9/223/7  & 0/232/0  & 0/232/3 & 0/232/13  & 23/209/10 & 118/114/9  & 0/232/0 &  225/7/27  \\
~ & recall(\%) & 0.43 & 3.9 & 0 & 0 & 0 & 9.9 & 50.9 & 0 & 97.0 \\
~ & precision(\%) & 100 & 56.25 & N/A & 0 & 0 & 69.7 & 92.9 & N/A & 86.3  \\
\bottomrule
\end{tabular}
\end{table*}
    
We analyze the experimental results, \emph{SolidityCheck} performs best in \emph{recall}, followed by \emph{SmartCheck}. \emph{SmartCheck} and \emph{SolidityCheck} are essentially static analysis approaches based on source codes.
The reason for the difference in \emph{recall} rate is that \emph{SolidityCheck}'s problem detection criteria are more accurate and reasonable, and consequently \emph{SmartCheck} may miss some statements that \emph{SolidityCheck} think are problematic. 
In terms of \emph{precision}, \emph{SmartCheck} is the best (\emph{ContractFuzzer} reports too few problems), followed by \emph{SolidityCheck}. 
The reason for the low accuracy of \emph{SolidityCheck} is that \emph{SolidityCheck}'s checking strategy is biased towards \emph{recall}. \emph{SolidityCheck} reports some statements with vague characteristics of the problems, so there are relatively more misjudgements. In general, we think that the higher \emph{recall} rate brings more benefits than the lower \emph{precision} rate, and the gap between \emph{SolidityCheck} and \emph{SmartCheck} is not obvious (6.6\%).
Consequently, we think that the overall detection quality of \emph{SolidityCheck} is better.

\subsubsection{Important Vulnerabilities}
To answer RQ4, we design a set of experiments to detect some important vulnerabilities, such as re-entrancy vulnerability and integer overflow problems. Because \emph{SolidityCheck} uses program instrumentation to detect these two problems while the other contrast tools only scan codes, to unify evaluation indicator, we use the following criteria for the ``effectiveness" of different tools: 
\begin{itemize}
    \item For \emph{SolidityCheck}, inserted codes effectively prevent problems from occurring. 
    \item For other tools, the correct location of the problem was reported.
\end{itemize}

The first experiment is performed for the re-entrancy vulnerability problem. We used Reentrancy.sol of \emph{not-so-smart-contracts}~\cite{NotSoSmartContracts} and Reentrance.sol of \emph{Ethernaut}~\cite{Ethernaut} as test cases, both of which are representative and problematic smart contracts.
\emph{Y} represents the effective response of the tool, and \emph{N} represents not, and we use \emph{N/A} to flag a tool to analyze the contract failure. The experimental results are shown in Table~\ref{Table9}.

\begin{table*}[h]
\centering
\caption{Experiment results of re-entrancy vulnerability}
\label{Table9}
\begin{tabular}{c|cccccccccc}
\toprule
  & \emph{Solhint}   & \emph{SmartCheck} & \emph{Mythx} & \emph{Oyente} & \emph{Osiris} & \emph{Securify}  & \emph{ContractFuzzer} & \emph{Remix} & \emph{SolidityCheck} \\
\midrule
Reentrancy.sol & N  & N  & N & Y & Y & Y & Y & Y & Y \\
\midrule
Reentrance.sol & N    & N    & N & N/A & N/A & N/A & Y & Y & Y \\
\bottomrule
\end{tabular}
\end{table*}

It is noteworthy that \emph{Securify} reports that all statements that use \emph{call} instruction to send ethers may introduce re-entrancy problems. \emph{Remix} reports that all transfer statements that do not comply with the \emph{CEI} (Checks-Effects-Interaction) mode may introduce re-entrancy vulnerabilities, even if the statement uses \emph{transfer} instructions to send ethers.

In the second experiment, we use the integer\_overflow\_1.sol (Abbreviation:Interger.sol) contract of \emph{not-so-smart-contracts} project~\cite{NotSoSmartContracts}, Token.sol of \emph{Ethernaut}~\cite{Ethernaut} and the BEC contract (Abbreviation:BEC.sol) source code~\cite{BecToken} as test cases. BEC contract is the smart contract that has lost the most because of the integer overflow problem so far. The experimental results are shown in Table~\ref{Table10}. \emph{Y} represents the effective response of the tool, while \emph{N} represents not, and we use \emph{N/A} to indicate that a tool occurs a failure when the contract is analyzed.

\begin{table*}[h]
\centering
\caption{Experimental results of integer overflow problem}
\label{Table10}
\begin{tabular}{c|ccccccccc}
\toprule
& \emph{Solhint} & \emph{ContractFuzzer}  & \emph{Securify} & \emph{Remix} & \emph{SmartCheck} & \emph{Mythx} & \emph{Oyente} & \emph{SolidityCheck}  & \emph{Osiris}   \\
\midrule
BEC.sol  & N  & N & N & N & N &N & N& Y  & Y \\
\midrule
Integer.sol & N & N &N&N&N&Y& Y &Y & Y  \\
\midrule
Token.sol  & N & N & N & N & N & Y& Y & Y  & Y \\ 
\bottomrule
\end{tabular}
\end{table*}

It can be seen from the experimental results. For re-entrancy vulnerabilities problem, \emph{Securify}, \emph{Remix}, and \emph{SolidityCheck} have responded effectively. For integer overflow problems, only \emph{SolidityCheck} effectively handles all integer overflows and \emph{Mythx} misses one. Listing 22 shows the codes inserted into the BEC contract by \emph{SolidityCheck}. The codes in line 3 inserted by \emph{SolidityCheck} effectively prevents the integer overflow problem in line 2, which is the exact problem statement that causes the attack of the BEC contract.
\begin{lstlisting}[     language=C++,
                            breaklines=true,
                            title={Listing 22: the part of codes after inserting code into the BEC contract},
                            captionpos=bc,
                            numbers=left,
                            basicstyle=\footnotesize,
                            numbersep=1pt, keywordstyle=\color{blue},
                            commentstyle=\color{blue}
                            ]
    uint cnt = _receivers.length;
    uint256 amount = uint256(cnt) * _value;
    require(uint256(cnt)==0 || amount/uint256(cnt)==_value);
    \end{lstlisting}
    
\subsubsection{Answers to Research Questions}
We designed a set of experiments to obtain the answers for the four research questions raised in Section~\ref{Sec_Experimental_questions}. Now, based on our experimental results, we give corresponding answers to these questions.
\begin{itemize}
    \item \emph{Answer1}: \emph{SolidityCheck} can generate detection results for any smart contract developed in \emph{Solidity} language without restricting the programming style and lines of codes of the contract under test.
    \item \emph{Answer2}: \emph{SolidityCheck} spends the least time among all tools. In comparison with \emph{SmartCheck}, the detection efficiency of \emph{SolidityCheck} is also significantly better than that of \emph{SmartCheck}.
    \item \emph{Answer3}: According to the problematic statement standard defined by us, \emph{SolidityCheck} is significantly better than the contrast tools in \emph{recall} rate, while \emph{SolidityCheck} lags behind \emph{SmartCheck} in \emph{precision} rate only a little. But for smart contracts, \emph{recall} is far more important than \emph{precision}, and we think that the overall detection quality of \emph{SolidityCheck} is better than similar tools.
    \item \emph{Answer4}: In our experiments, \emph{SolidityCheck} is the only tool that can resist both re-entrancy vulnerabilities and integer overflow problems.
\end{itemize}

\section{Discussion}
\label{sec_disucssion}
Source code analysis pursues detection efficiency and problem recall rate. Our desire is to bring better user experience and more complete security for smart contract developers.
However, \emph{SolidityCheck} still has its own shortcoming in its current state. We summarize the existing problems and the corresponding solutions in the following:
\begin{enumerate}
    \item \emph{SolidityCheck} cannot feedback detection information, which makes developers have to use the editor again to modify the problematic statement after the detect results are gotten, which brings bad user experience. We plan to make the feedback of \emph{SolidityCheck}'s detect results timely and in a more convenient way in the future.
    \item The rapid development of smart contract field has brought endless security accidents and a large number of related literature. Many smart contract problems that we do not find at the beginning are discovered now. Based on our capabilities, we cannot always keep updating our target problems set while studying \emph{SolidityCheck}, which makes it impossible for \emph{SolidityCheck} to detect some smart contract problems. We plan to continue updating \emph{SolidityCheck} in the future so that it can detect more types of problems.
    \item Source code based analysis can bring a good problem \emph{recall} rate, which is very important for smart contracts. However, this does not mean that \emph{accuracy} rate should be ignored, the low accuracy rate may make the results of \emph{SolidityCheck} not instructive. In the following study, we will provide more accurate regular expressions, more technical means to continuously enhance the \emph{accuracy} of \emph{SolidityCheck}.
    \item Our re-entrancy vulnerability prevention approach can address the most dangerous type of re-entrany vulnerability (re-entrancy  vulnerability  with  ether  transfer), but it does not mean that another type of re-entrancy vulnerability (re-entrancy vulnerability with no ether  transfer) can be ignored. We plan to further study new approach to detect or prevent this type of re-entrancy vulnerability in the future.
\end{enumerate}

\section{Related work}\label{sec_Related work}
\subsection{Safety status of smart contracts in Ethereum}
Some researchers focus on investigating on the security status of Ethereum smart contracts. According to their study, we can understand the worrying security status of Ethereum smart contracts. Destefanis et al.~\cite{destefanis2018smart} investigated the freezing accident of parity wallet in Ethereum, put forward the smart contract programming mode which should be avoided centrally, and finally put forward the necessity of building block chain software engineering. Atzei et al.~\cite{atzei2017survey} analyzed the academic literature, blogs and forums in the field of smart contract security in Ethereum, combined with their programming experience, expounded the vulnerabilities of Ethereum and its mainstream smart contract programming language (Solidity), and proposed the classification of common programming vulnerabilities that may lead to vulnerabilities. 
Nikoli\'c et al.~\cite{nikolic2018finding} described vulnerabilities in smart contracts as traceable attributes, and tried to find greedy (locked money), prodigal (which may leak ethers to any user) and suicidal (which can be killed by anyone) smart contracts through cross-contract symbolic analysis and verification. They implemented \emph{MAIAN}, and with this tool, they succeeded in finding vulnerabilities in parity wallet. Wang et al.~\cite{wang2019blockchain} proposed a research framework for smart contracts based on a six-layer architecture and described the problems existing in smart contracts in terms of contract vulnerability, limitations of the blockchain, privacy, and law. Through interviews with smart contract developers, Zou et al.~\cite{8847638} revealed that smart contract developers still face many challenges when developing contracts, such as rudimentary development tools, limited programming languages and Ethereum virtual machines, and difficulties in dealing with performance issues. However, in our opinion, these studies do not propose an effective classification criterion for smart contract problems, a general classification criterion for \emph{Solidity} language is still lacking.
\subsection{Ethereum virtual machine bytecodes}
Analyzing the security of smart contracts based on the bytecodes of the EVM (Ethereum virtual machine) is the mainstream methods at present. Grishchenko et al.~\cite{grishchenko2018semantic} proposed the first small semantics of EVM by formalizing bytecodes, obtained executable codes, and successfully verified the official EVM test suite. Luu et al.~\cite{luu2016making} used attribute analysis to verify the security of smart contracts and developed \emph{Oyente} to detect security vulnerabilities in smart contracts. Albert et al.~\cite{albert2019safevm} introduced the advanced validation engine used to validate the C language program into the security validation of smart contracts, validated the security of contracts through the C language program validation engine, and finally output a report with validation results. Their work bridges the gap between today's advanced C-program verification technology and the security of Ethereum smart contracts. Tann et al.~\cite{tann2018towards} used machine learning technology to detect security vulnerabilities in smart contracts. The approach uses LSTM (Long Short-Term Memory) network to learn the behavior pattern of smart contract bytecodes, and thus obtains better detection accuracy than the \emph{MAIAN} tool developed by Nikoli'c et al.~\cite{nikolic2018finding}. Torres et al.~\cite{torres2018osiris} realized the integer defect detection tool \emph{Osiris} of Ethereum smart contract through symbol execution and pollution analysis. The tool can detect arithmetic errors, truncation errors, and signature errors. As far as we know, this tool is one of the few Ethereum smart contract analysis tools that can handle integer errors. Tsankov et al.~\cite{tsankov2018securify} analyzed the bytecodes of Ethereum virtual machine, obtained the exact semantics of the statement, and judged the security of smart contracts. Based on semantics, a tool called \emph{Securify} is realized, which can scan the security of Ethereum smart contract. Chen et al.~\cite{chen20171} focused on the execution cost of Ethereum smart contracts. Through investigation and analysis, they found that many recommended smart contract compilers generate bytecodes that contain expensive patterns, even though these bytecodes have been optimized. To this end, they proposed and developed \emph{Gasper}, a symbol-based execution tool for detecting expensive patterns in bytecodes.
Bragagnolo et al.~\cite{bragagnolo2018smartinspect} realized \emph{SmartInspect}, a smart contract debugging tool, by analyzing the bytecode distribution in the Ethereum virtual machine memory.

Various analysis methods based on the bytecodes of the Ethereum virtual machine can accurately detect security vulnerabilities in contracts. However, the detection efficiency of them is relatively low, and they do not adapt quickly to Ethereum updates.
Furthermore, these methods cannot accurately locate the possible issues in the source codes of smart contracts.
\subsection{Smart contract static code analysis}
Static code analysis based on contract source codes has high problem coverage and detection efficiency, and can also detect problems that affect the readability of the codes. It is a useful supplement to the methods based on EVM bytecode. To the best of our knowledge, only Tikhomirov et al.~\cite{tikhomirov2018smartcheck} have done some research on static code analysis based on the source codes of the Ethereum smart contract. They classify and summarize the existing Ethereum smart contract problems, and use lexical analysis, grammar analysis and other technologies to achieve static code analysis of smart contracts. However, their problem detection criteria are not very accurate, and cannot detect important security problems such as integer overflow problems and re-entrancy vulnerabilities which have a significant impact on the security of smart contracts.


\section{Conclusion and  future work} \label{sec_conclusion}
With the vigorous development of blockchain technology, the Ethereum smart contract has been paid more and more attention. In this paper, we propose a novel approach, namely \emph{SolidityCheck} to detect Ethereum smart contract problems based on regular expressions and program instrumentation. 
A series of experiments show that the tool corresponding to \emph{SolidityCheck} has more advantages than existing ones in terms of detection quality and  efficiency, and \emph{SolidityCheck} is can deal with important issues of smart contracts such as re-entrancy vulnerabilities and integer overflow problems.

For future work, we have the following plans:
\begin{itemize}
\item Determining a comprehensive, reasonable, accurate and up-to-date code problem criterion for Ethereum smart contract.
\item According to the criterion of Ethereum smart contract problems, the problems that cannot be detected by \emph{SolidityCheck} can be added to \emph{SolidityCheck} in the future.
\item Continuously optimizing \emph{SolidityCheck} to further improve detection quality, performance, and stability.
\end{itemize}

\section{Acknowledgements}
The work is supported by the National Natural Science Foundation
of China under Grant No. 61572171, the Natural Science Foundation of Jiangsu Province under Grant No. BK20191297, and the Fundamental
Research Funds for the Central Universities under Grant No. 2019B15414. 
\bibliographystyle{IEEEtran}
\bibliography{bibfile2}

\begin{IEEEbiography}[{\includegraphics[width=1in,height=1.25in,clip,keepaspectratio]{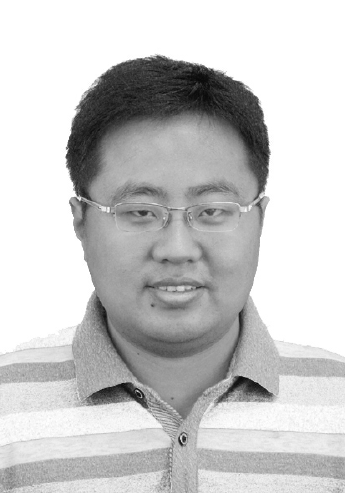}}]{Pengcheng Zhang}
received the Ph.D. degree in computer science from Southeast University in 2010. He is currently an associate professor in College of Computer and Information, Hohai University, Nanjing, China, and was a visiting scholar at San Jose State University, USA. His research interests include software engineering, service computing and data mining. He has published in premiere or famous computer science journals. He was the co-chair of IEEE AI Testing 2019 conference. He served as technical program committee member on various international conferences. He is a memeber of the IEEE.
\end{IEEEbiography}
\vspace{-14 mm}
\begin{IEEEbiography}[{\includegraphics[width=1in,height=1.25in,clip,keepaspectratio]{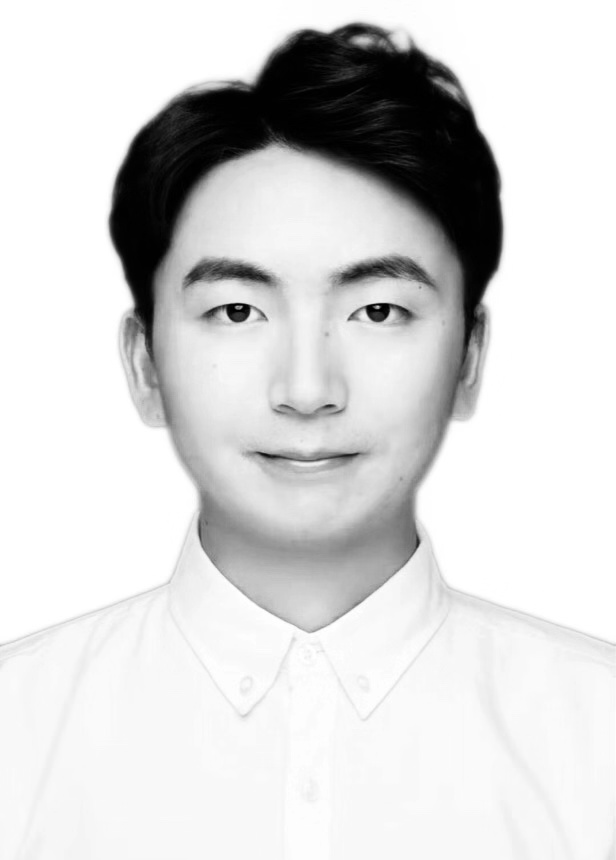}}]{Feng Xiao} received the bachelor's degree in computer science and technology from Hohai university in 2019. He is currently working toward the M.S. degree with the College of Computer and Information, Hohai University, Nanjing, China. His current research interests include smart contract security and software engineering.
\end{IEEEbiography}
\vspace{-14 mm}
\begin{IEEEbiography}[{\includegraphics[width=1in,height=1.25in,clip,keepaspectratio]{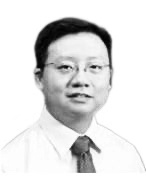}}]{Xiapu Luo} is an assistant professor with the Department of Computing and an Associate Researcher with the Shenzhen Research Institute,
The Hong Kong Polytechnic University. He received the Ph.D. degree in Computer Science from The Hong Kong Polytechnic University, and was a Post-Doctoral Research Fellow with the Georgia Institute of Technology. His research focuses on smartphone security and privacy, network security and privacy, and Internet measurement.
\end{IEEEbiography}
\vspace{-14 mm}


 \newpage 
\begin{appendices}
\section{Problems and their corresponding regular expression}\label{Appendix_B}
We introduce the matching rules for \emph{SolidityCheck} here:
\begin{itemize}
    \item \emph{Balance equality}. \emph{SolidityCheck} matches the statement with the characteristics of formula 2.1.
    \item \emph{Mishandled exceptions}. \emph{SolidityCheck} matches the statement with the characteristics of formula 2.2.
    \item \emph{DoS by external contract}. \emph{SolidityCheck} matches statements with characteristics of formula 2.3 or 2.4.
    \item \emph{Using tx.origin for authentication}. \emph{SolidityCheck} matches the statement with 2.5 characteristics.
    \item \emph{Missing constructor}. \emph{SolidityCheck} detects whether the contract contains statements with 2.6 or 2.7 characteristics, and reports the line number of the contract header if no.
    \item \emph{Locked money}. \emph{SolidityCheck} first detects whether the contract contains statements with the 2.8 characteristics. If so, \emph{SolidityCheck} detects statements with 2.9 or 2.10 characteristics in the non-inline assembly code, and detects statements with 2.11 characteristics in the inline assembly code, if such statements do not exist (2.9, 2.10, 2.11), \emph{SolidityCheck} reports this problem.
    \item \emph{Unsafe type inference}. \emph{SolidityCheck} detects the statement with the 2.12 characteristics and requires the number to the right of the "=" to be less than 2$^198$ and greater than (-2)$^{197}$. Because of beyond these two boundary values, variables are matched to the type in \emph{Solidity} that has the largest storage space for shaping.
    \item \emph{byte\lbrack\ \rbrack}. \emph{SolidityCheck} matches the statement with 2.13 characteristics.
    \item \emph{Costly loop}. \emph{SolidityCheck} detects statements with the  2.14-2.19 characteristics, these formulas define loops that contain identifiers or function calls in the conditional judgment part of a for-statement or while-statement. In addition, loops with a maximum execution of more than 23 sentences will also be considered costly loops (we explain why in appendix~\ref{Appendix_C}).
    \item \emph{Timestamp dependence}. \emph{SolidityCheck} matches the statement with 2.20 characteristics.
    \item \emph{Token API violation}. In our detection, firstly, the regular expression 2.21 matches the contract or interface whose name contains ERC20, ERC721, and ERC165. If there is a statement that conforms to the above characteristics, the contract is considered to be developed by following one of ERC20, ERC721 or ERC165, and the contract or interface of the token standard is declared as ERC contract. Then, in all contract codes directly or indirectly inherited from the ERC contract, the pattern detects whether there are any exception-causing codes (using require, assert, throw, revert()) in functions with type 2.22 characteristics. If there is, it is identified as a problem; if not, there is no problem.
    \item \emph{Using fixed point number type}. \emph{SolidityCheck}  matches  the statement with the characteristics of formula 2.23.
    \item \emph{Private modifier}. \emph{SolidityCheck} detects statements with 2.24 characteristics but without 2.25 characteristics.
    \item \emph{Redundant refusal of payment}. When the compiler version is higher than 0.4.0, regular expression 2.26 represents the fallback function that receives an external payment and we detect in the body of the function whether to use \emph{revert}() or \emph{throw}. If so, there is a problem.
    \item \emph{Compiler version problem}. \emph{SolidityCheck} detects statements that have characteristics of type 2.27, or statements that have characteristics of type 2.28 but do not contain the "$\textless$" symbol. In this problem, to promote the popularity of the newer version of security specifications, the pattern retrieves whether the contract contains statements with 2.29 characteristics. If not, it’s considered a problem.
    \item \emph{Style guide violation}. \emph{SolidityCheck} detection has one of the characteristics of formula 2.30, 2.31 and 2.32.
    \item \emph{Integer division}. \emph{SolidityCheck}  matches  the  statement with the characteristics of formula 2.33.
    \item \emph{Implicit visibility level}. \emph{SolidityCheck} detects statements that have one of the characteristics of 2.34 to 2.40, but do not have the characteristics of type 2.41.
\end{itemize}
    Please refer to the regular expressions in the table on the page 21 for detail.

\section{More precise ``costly loop" problem}\label{Appendix_C}    
Tikhomirov et al.\cite{tikhomirov2018smartcheck} proposed the definition of \emph{costly loop} problems, and proposed the corresponding detection criteria for ``costly loop" problems:
\begin{itemize}
\item ``for"-statements with function calls or identifiers in the conditional part.
\item ``while"-statements with function calls in conditional part.
\end{itemize}
In other words, the statements shown in Listing 1 are not considered as a costly loop. But the loop will have 20000 statements executed. In fact, such a loop must be expensive, and, likely, it will not be packaged into any block.
          \begin{lstlisting}[     language=C++,
                            breaklines=true,
                            captionpos=bc,
                            basicstyle=\footnotesize,
                            keywordstyle=\color{red},
                           commentstyle=\color{blue},
                            title={Listing 1: costly loop},
                            stepnumber=1
                            ]
    for (int i = 0; i < 10000; i++){
        if (msg.sender == receiver1)
            msg.sender.transfer(1 wei);
    }
    \end{lstlisting}


The reason for these omissions is that Tikhomirov et al's criteria~\cite{tikhomirov2018smartcheck} do not take into account that the maximum number of loops is a fixed value. To reduce such omissions, we need to address the following two issues:
\begin{enumerate}
    \item In current Ethereum, how much gas spent can be considered an expensive transaction?
    \item If all statements executed by a contract call are in a loop, how many statements executed by the loop will make the call an expensive transaction?
\end{enumerate}

To get the answers for these two questions, we need to obtain the following three values:
\begin{enumerate}
    \item How much gas consumed can be called an expensive transaction?
    \item How much gas does a statement consume on average in a loop?
    \item How to get the maximum execution statement number of a loop?
\end{enumerate}

To this end, we carry out the following statistical analysis.

Let,
\begin{itemize}
    \item the total daily gas consumption be \emph{G},
    \item the gas consumption per transaction be \emph{g},
    \item the number of transactions per day be \emph{n}.
\end{itemize}
   Then \emph{G}, \emph{g} and \emph{N} satisfy the following formulas:
        \makeatletter
    \@addtoreset{equation}{section}
    \makeatother
    \renewcommand{\theequation}{\arabic{section}.\arabic{equation}}
    \begin{equation}
    G = \sum_{i=1}^ng
    \end{equation}
The average gas consumed per transaction per day \emph{AG} satisfies the following formulas:
    \makeatletter
    \@addtoreset{equation}{section}
    \makeatother
    \renewcommand{\theequation}{\arabic{section}.\arabic{equation}}
    \begin{equation}
    AG = G / n
    \end{equation}
    
\begin{figure}[h]
    \centering
    \includegraphics[width=8cm,height=5.8cm]{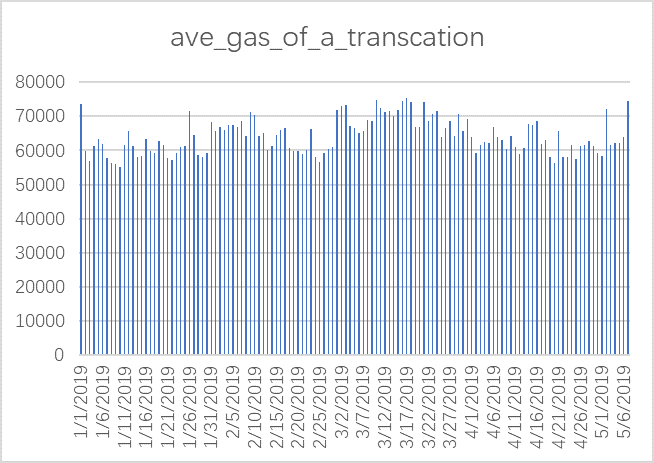}
    \caption{Average gas consumption per transaction per day}
    \label{fig21}
    \end{figure}

By collecting daily gas consumption and daily trading volume from Ethereum from January 1, 2019, to May 7, 2019, we get the value of \emph{AG}, as shown in Fig.~\ref{fig21}.

\begin{figure}[h]
    \centering
    \includegraphics[scale=0.45]{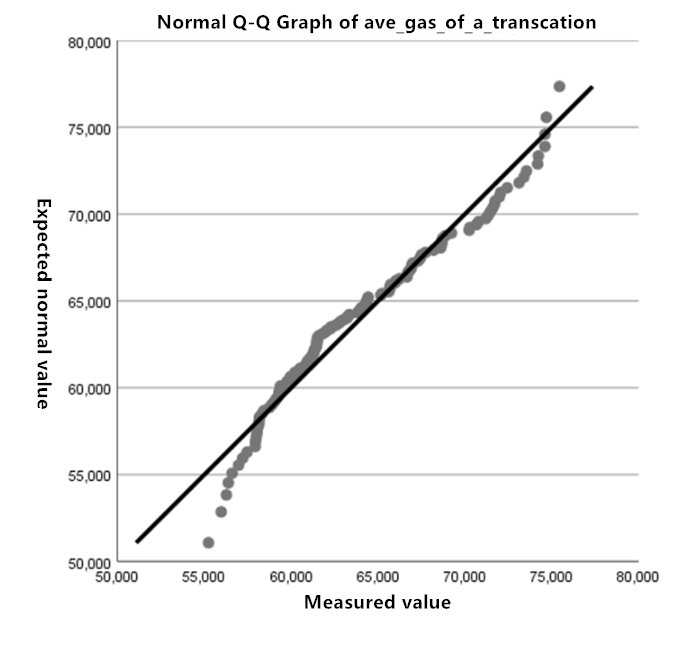}
    \caption{Normal Q-Q Diagram Generated by SPSS}
    \label{fig22}
    \end{figure}

We use SPSS~\cite{george2011spss} software to generate a normal Q-Q graph of \emph{AG} data. The result shows that the data points are evenly distributed on both sides of the normal distribution line. Therefore, it is considered that \emph{AG} data conform to normal distribution. The analysis results are shown in Fig.~\ref{fig22}.

    According to the nature of normal distribution,
    let:
    \begin{itemize}
        \item the \emph{expectation} be \emph{E}.
        \item the \emph{standard deviation} be \emph{D}.
    \end{itemize}
     The transactions consuming gases less than \emph{(E+D)} account for 84.3\% of the total transactions. 
  
    \begin{table}[h]
    \centering
    \caption{Partial information on AG values}
    \label{Tab_Partial}
\begin{tabular*}{6cm}{p{3cm}<{\centering}|p{3cm}<{\centering}}
\toprule
Sample size & 127 \\
\midrule
Average value & 64212.5996 \\
\midrule
Standard deviation & 5087.94154 \\
\midrule
Minimum  & 55238.50 \\
\midrule
Maximum  & 75442.82 \\
\bottomrule
\end{tabular*}
\end{table}

A transaction that consumes gases more than the sum of an \emph{expectation} and a \emph{standard deviation} is set as an expensive transaction. The reason is that 84.3\% of the transactions consumed gas below that value.

Table~\ref{Tab_Partial} contains the analysis information of \emph{AG} data. According to the analysis data, the criterion for expensive transactions is the consumption of more than 69301 gases.

The number of gases consumed by various statements is specified in detail in~\cite{EthereumYellowPaper}.

According to the possibility of different kinds of sentences appearing in a loop, this paper divides 36 kinds of sentences into three levels, namely:
    \begin{itemize}
    \item Types of statements frequently used in loops
    \item Types of statements occasionally used in loops
    \item Types of statements that are not possible to use in loops
    \end{itemize}
    
We divide 36 statements into three categories and the results are shown in Table~\ref{thirtySix}.

    \begin{table}
    \centering
    \caption{Classification of Frequent Use of 36 Sentences in Loops}
    \label{thirtySix}
    \begin{tabular*}{7cm}{p{2cm}<{\centering}p{2cm}<{\centering}p{2cm}<{\centering}}
    \toprule
    Frequently & Occasionally & Impossible\\
    \midrule
    14 & 17 & 5 \\
    \bottomrule
    \end{tabular*}
    \end{table}

Based on this table, we get the following formula by setting the weight of frequently used sentences to $x$ and the weight of occasionally used sentences to $y$.
        \makeatletter
    \@addtoreset{equation}{section}
    \makeatother
    \renewcommand{\theequation}{\arabic{section}.\arabic{equation}}
    \begin{equation}
    14x + 17y = 1\label{4_9}
    \end{equation}
        \makeatletter
    \@addtoreset{equation}{section}
    \makeatother
    \renewcommand{\theequation}{\arabic{section}.\arabic{equation}}
    \begin{equation}
    x > y\label{4_10}
    \end{equation}
    For the solutions of (\ref{4_9}) (\ref{4_10}),
    we choose a reasonable solution from multiple solutions:\\
    \[
    x=0.052,y=0.016
    \]
    According to the number of gases consumed by different kinds of statements and the weight of different statements. Let,
    \begin{itemize}
        \item \emph{ave\_gas} be the average gas consumed by each statement in a loop,
        \item \emph{w} be the weight of each statement,
        \item \emph{gas} be the number of gases consumed by each statement.
    \end{itemize}
      Then we get following formula:
      \makeatletter
    \@addtoreset{equation}{section}
    \makeatother
    \renewcommand{\theequation}{\arabic{section}.\arabic{equation}}
    \begin{equation}
    ave\_gas = \sum_{n=1}^{36}{gas}*w
    \end{equation}
    According to the above formula, the average gas consumption per statement in a loop is 2928 gas.

    Let,
    \begin{itemize}
        \item A loop executes at most $L_{max}$ statements.
        \item A loop runs at most \emph{C} times.
        \item A loop executes up to \emph{L} statements at a time.
    \end{itemize}
     The relationship between the above three is as follows:
      \makeatletter
    \@addtoreset{equation}{section}
    \makeatother
    \renewcommand{\theequation}{\arabic{section}.\arabic{equation}}
    \begin{equation}
    L_{max} = C * L
    \end{equation}
    By definition, the maximum number of statements executed by a loop is 24 in Listing 2.
    
        \begin{lstlisting}[     language=C++,
                            breaklines=true,
                            title={Listing 2: A loop that executes 24 statements},
                            captionpos=bc,
                            basicstyle=\footnotesize,
                            keywordstyle=\color{blue},
                            commentstyle=\color{blue}
                            ]
    for (int i = 0; i < 12; i++){
        my_balance += 1;
        sender_balance -= 1;
    }
\end{lstlisting}
    
    We assume that if all statements executed by a call are statements in a loop, and the loop header statements execution does not consume gas, a loop that conforms to the following formula will be judged as a costly loop. 
    Let,
    \begin{itemize}
        \item the gas standard of expensive transaction consumption be \emph{C$_{exp}$}.
        \item the maximum number of statements executed in a loop be \emph{L$_{max}$}.
        \item the average gas consumption of each statement be \emph{G$_{ave}$}.
    \end{itemize}
     So the formula is as follows:
          \makeatletter
    \@addtoreset{equation}{section}
    \makeatother
    \renewcommand{\theequation}{\arabic{section}.\arabic{equation}}
    \begin{equation}
    G_{ave} * L_{max} \geq C_{exp}
    \end{equation}
  Based on the above calculation, when the maximum number of statements executed in a loop is more than 23, it will be judged as a costly loop (The standard is adjustable).

\begin{table*}[h]
\centering
\label{Appendix_B}
\begin{tabular*}{18cm}{|p{2cm}|p{13.5cm}|p{1.1cm}<{\centering}|}
\toprule
Problem name & Regular expressions & Number \\
\midrule
 Balance equality &  $   \wedge(\setminus s)((if)|(while)|(require))(\setminus s)* (\setminus ()(.)* (((this. balance)(\setminus s)*(==) (\setminus s)*(\setminus d))+(ether))|((\setminus d)+(\setminus s)* (ether)(\setminus s)*(==)(\setminus s)* (this.balance ))) (.)*(\setminus ))(.)*\$ $ & $(2.1)$
    \\
 \midrule
Mishandled exceptions & $ (\wedge(\setminus s)*((if)|(require))(\setminus s)*(\setminus() (.)*((\setminus w)|(\setminus ()|(\setminus ))|(\setminus [)|  (\setminus ])|(.)) (( send)|(delegatecall)|(call)|(callcode))(.)* (\setminus()(.)*(\setminus))(.)*(\setminus)))$ & $(2.2)$
 \\
 \midrule
 \multirow{2}{2cm}{Dos by external contract} &    $((if)|(require))(\setminus s)*(\setminus ()(.)*(\setminus .)(\setminus w)+(\setminus ()(.)*(\setminus ))(.)*(\setminus )) $ & $(2.3)$ \\
  ~ & $   (for)(\setminus s)*(\setminus ()(.)*(;)(.)+(\setminus .)(\setminus w)+(\setminus ()(.)*(\setminus ))(.)*(;)(.)*(\setminus ))$ & $(2.4)$ \\

  \midrule
  Using \emph{tx.origin} for authentication & $\wedge (\setminus s)*((require)|(if))(\setminus s)*(\setminus ()(.)*(tx.origin)(.)*(\setminus ))$ & $(2.5)$ \\
  \midrule
  \multirow{2}{2cm}{Missing constructor} & $\wedge (\setminus s)*(constructor)(\setminus s)*(\setminus ()$ & $(2.6)$ \\
~ & $\wedge(\setminus s)*(function)(\setminus s)*(\emph{contractName})(\setminus s)*(\setminus ()$ & $(2.7)$ \\
  
  \midrule

 \multirow{4}{2cm}{Locked money} & $  (\setminus s)*(function)(\setminus s)(.)+(\setminus ))(.)*(\setminus s)(payable)((\setminus s)|(\setminus \{)|(;))$  & $(2.8)$ \\
 ~ &$  (\setminus .)((transfer)|(send))(\setminus s)*(\setminus ()(.)+(\setminus ))$ & $(2.9)$ \\
 ~ & $    (\setminus .)(call)(\setminus .)(\setminus s)*((value)|(gas)(\setminus (.)+(\setminus ))(\setminus .)(value)))(\setminus ()$ & $(2.10)$ \\
 ~ & $ (\setminus b)((delegatecall)|(staticall)|(callvalue)|(call))(\setminus s)*(\setminus ()$ & $(2.11)$ \\
\midrule
Unsafe type inference & $ (\setminus b)(var)(\setminus b)(\setminus s)+(\setminus w)+(\setminus s)*(=)(\setminus s)*(\setminus d)+(\setminus b)$ & $(2.12)$ \\
\midrule

 byte\lbrack\ \rbrack & $(\setminus s)*(byte)(\setminus s)*(\setminus [)(\setminus s)*(\setminus ])(\setminus s)$ & $(2.13)$ \\
\midrule
\multirow{6}{2cm}{Costly loop} & $   (\setminus b)(for)(\setminus s)*(\setminus ()(.)*(;)(.)*(\setminus .)(.)*(;)(.)*(\setminus ))$ & $(2.14)$ \\
~ & $ (\setminus b)(while)(\setminus s)*(\setminus ()(.)*(\setminus .)(.)*(\setminus ))$ & $(2.15)$ \\
~ & $  (\setminus b)(for)(\setminus s)*(\setminus ()(.)*(;)(.)*(\setminus w)+(.)*(;)(.)*(\setminus ))$ & $(2.16)$ \\
~ & $ (\setminus b)(while)(\setminus s)*(\setminus ()(.)*(\setminus w)+(.)*(\setminus ))$ & $(2.17)$ \\
~ & $  (\setminus b)(for)(\setminus s)*(\setminus ()(.)*(;)(.)*(\setminus ()(.)*(\setminus ))(.)*(;)(.)*(\setminus ))$ & $(2.18)$ \\
~ & $ (\setminus b)(while)(\setminus s)*(\setminus ()(.)*(\setminus ()(.)*(\setminus ))(.)*(\setminus ))$ & $(2.19)$ \\
  \midrule
  Timestamp dependence & $(((\setminus b)(now)(\setminus b))|((\setminus b)(block.timestamp)(\setminus b)))$  & $(2.20)$ \\

\midrule
\multirow{4}{2cm}{Token API violation} & $ (\setminus s)*((contract)|(interface))(\setminus s)+(\setminus w)*((ERC)|(ERc) |(eRC)|(eRc)| (ErC)|(Erc)|(erC)|(erc))((20)|$ & ~ \\
~ & $(721)|(165))(\setminus w)*((\setminus s)|(\{)|(;))$ & $(2.21)$ \\
~ & $(\setminus s)*(function)(\setminus s)+(\setminus b)((transfer)|(transferFrom)|(arrrove)|(supportsInterface)|(isApprovedFo$ & ~ \\
~ & $rAll))(\setminus b)$ & $(2.22)$ \\
\midrule
 Using fixed point number type & $(\setminus b)((unfixed)(fixed))((\setminus d)\{1,3\}(x)(\setminus d) \{0,2\})?(\setminus s)+(\setminus w)+$ & $(2.23)$ \\
\midrule
  \multirow{2}{2cm}{Private modifier} & $    (\setminus b)(private)(\setminus b)$ & $(2.24)$ \\
  ~ & $    (\setminus b)(fcuntion)(\setminus b)$ & $(2.25)$ \\
  \midrule
  Redundant refusal of payment & $ (\setminus b)(function)(\setminus s)*(\setminus ()(\setminus s)*(\setminus ))(.)*(\setminus s)+(payable)(\setminus s)*$ & $(2.26)$ \\
\midrule
\multirow{3}{2cm}{Compiler version problem} & $ (\setminus s)*(pragma)(\setminus s)+(solidity)(\setminus s)+(\setminus \wedge)(\setminus d)(\setminus .)(\setminus d)(\setminus .)(\setminus d)(\setminus s)*(;)$ & $(2.27)$ \\
~ &$ (\setminus s)*(pragma)(\setminus s)+(solidity)(\setminus s)+(\setminus >)(\setminus =)(\setminus d)(\setminus .)(\setminus d)(\setminus .)(\setminus d)$& $(2.28)$ \\
~ & $ (\setminus s)*(pragma)(\setminus s)+(experimental)(\setminus s)+$ &  $(2.29)$ \\
\midrule
 \multirow{3}{2cm}{Style guide violation} & $(\setminus b)(function)(\setminus s)+[\wedge a-z](\setminus w)+$  & $(2.30)$ \\
 ~ & $ (\setminus s)*(event)(\setminus s)+[\wedge A-Z](\setminus w)+$ & $(2.31)$ \\
 ~ & $(\setminus b)(\setminus w)+(\setminus s)+(\setminus [)(.)*(\setminus ])$ & $(2.32)$ \\
 \midrule
   Integer division & $ (\setminus d)+(\setminus s)*(\setminus /)(\setminus s)*(\setminus d)+$ & $(2.33)$ \\
 \midrule
 \multirow{9}{2cm}{Implicit visibility level} & $  \wedge (\setminus s)*((uint)|(int))(\setminus d)\{0,3\}(\setminus s)+(\setminus w)+$ & $(2.34)$ \\
 ~ & $ \wedge (\setminus s)*((ufixed)|(fixed))((\setminus d)\{1,3\}(x)(\setminus d)\{0,2\})?(\setminus $ & $(2.35)$ \\
  ~ & $\wedge (\setminus s)*(bool)(\setminus s)+(\setminus w)+$ & $(2.36)$ \\
   ~ & $ \wedge (\setminus s)*(address)(\setminus s)+(\setminus w)+$ & $(2.37)$ \\
    ~ & $ \wedge (\setminus s)*(mapping)(\setminus s)*(\setminus ()(\setminus s)*(\setminus w)+(\setminus $ & $(2.38)$ \\
     ~ & $\wedge (\setminus s)*(((bytes)\{0,2\})|(byte))(\setminus s)+(\setminus w)+$ & $(2.39)$ \\
      ~ & $ \wedge (\setminus s)*(string)(\setminus s)+$ & $(2.40)$ \\
       ~ & $\wedge (\setminus s)*(\setminus w)+(\setminus s)*(\setminus [)(.)*(\setminus ])(\setminus s)+$ & $(2.41)$ \\ 
\bottomrule
\end{tabular*}
\end{table*}

 \section{SolidityCheck}\label{Appendix_A}
 According to the theoretical description, we implement \emph{SolidityCheck}, a static code analysis tool for Ethereum smart contracts written by C++. \emph{SolidityCheck} first formats the codes, and then detects the formatted code. \emph{SolidityCheck} consists of several classes, each of which is responsible for detecting or preventing only one problem. For each statement, each class filters keywords first, and if it contains a specific keyword, It matches statements (we use the \emph{regex} library in the C++ standard library to match). For statements with the problem characteristics described by regular expressions, \emph{SolidityCheck} records the line number of the problem statement (Line numbers refer to the formatted code file).

For some problems that cannot be judged by one line of statement, we use some programming tricks to make the detection ability of regular expressions span multiple lines. For example, in the case of \emph{costly loop}, we use bracket matching to get the maximum number of statements that are executed once by a loop, while regular expressions are used to detect \emph{for}-statements or \emph{while}-statements that may be problematic. Of course, most of the problem types detected by \emph{SolidityCheck} can be detected in one line after code formatting, which is why we use regular expressions and determine code formatting standards. For some problems involving several statements, we do use some programming tricks to enable \emph{SolidityCheck}'s detection capability to span multiple lines.

In our experiments, \emph{SolidityCheck} can cover nearly every sentence that we believe is problematic, which also proves the completeness of \emph{SolidityCheck}'s cross-statement detection ability.

To prevent problems, \emph{SolidityCheck} first retrieves statements that may introduce problems through regular expressions and then inserts code before and after statements through string splicing. Adding line breaks in string splicing will eventually achieve the effect of inserting codes before and after the problem statement in the output file.

\emph{SolidityCheck} provides seven functions, which are used from the CLI (command line interface). The functions and corresponding commands are as follows:
\begin{enumerate}
    \item  Getting help information ---- SolidityCheck -{}-help
\item Generating contracts to prevent re-entrancy vulnerabilities ---- SolidityCheck -{}-r
\item Generating contracts to prevent integer overflow problems ---- SolidityCheck -{}-o
\item Detecting of 18 other problems except integer overflow and re-entrancy vulnerabilities ---- SolidityCheck -{}-d
\item Adjusting costly loop standard ---- SolidityCheck -{}-g
\item Checking existing expensive cycling standards ---- SolidityCheck -{}-s
\item Batch testing ---- SolidityCheck -{}-f
\end{enumerate}

\end{appendices}
%

\end{document}